%% file: VLDB2026.tex
\documentclass[sigconf, nonacm, screen,pdfa]{acmart}
\input{setting_package.tex}

\begin{document}

\title[ReSequel: Robust LLM-assisted Query Rewriting and Optimization]{ReSequel: Robust LLM-assisted Query Rewriting and Optimization using Templatization and Sampling}

\author{Saeed Fathollahzadeh} 
\affiliation{%
	\institution{Concordia University}}
\email{saeed.fathollahzadeh@concordia.ca}

\author{Essam Mansour} 
\affiliation{%
	\institution{Concordia University}}
\email{essam.mansour@concordia.ca}

\author{Matthias Boehm} 
\affiliation{
    \institution{TU Berlin \& BIFOLD}}
\email{matthias.boehm@tu-berlin.de}

\input{0_Abstract}
\maketitle
\input{setting_paper.tex}

\input{1_Introduction}

\input{2_Background}
\input{3_Architecture}
\input{4_Templatization}

\input{5_Downsampling}

\input{6_Rewriting}

\input{7_Experiments}
\input{8_RelatedWork}

\input{9_Conclusions}

\balance
\bibliographystyle{ACM-Reference-Format}

\bibliography{References}

\end{document}

%% file: setting_package.tex
\usepackage{pgfplots}
\usepackage{algorithm}
\usepackage[noend]{algorithmic}
\usepackage{listings}
\usetikzlibrary{arrows.meta, chains, positioning, shapes.symbols}
\usetikzlibrary{decorations,calligraphy}
\usepackage{numprint}
\usepackage{subfigure}
\usepackage{mathtools}
\usepackage{amsmath,mleftright}
\usepackage{multirow}
\usepackage{pifont}
\usepackage{colortbl}
\usepackage{marginnote}

\newcommand{\eat}[1]{}
\usepackage{balance}
\usepackage{csquotes}

\usetikzlibrary{matrix,calc,positioning}
\usepackage{makecell}

\npstyleenglish

\def\shorten{\looseness=-1} 

\newcommand{\chline}{\arrayrulecolor{lightgray} \hline \arrayrulecolor{black}}
\definecolor{commentgreen}{RGB}{106,168,79}

\renewcommand{\algorithmiccomment}[1]{\hfill {\color{commentgreen}\textbf{\textit{// #1}}}}
\newcommand{\COMMENTLINE}[1]{\STATE {\color{commentgreen}\textbf{\textit{// #1}}}}

\newcommand{\p}{\ensuremath{\mathcal{P}}}
\newcommand{\qt}{\ensuremath{\mathcal{T}}}

\newcommand{\ts}{\ensuremath{\mathcal{S}}}
\newcommand{\R}{\ensuremath{\mathcal{R}}}
\newcommand{\D}{\ensuremath{\mathcal{D}}}

\newcommand{\Q}{\textbf{Q}}

\newcommand{\cat}{\ensuremath{\mathcal{C}}}
\newcommand{\QP}{\textbf{Q$^\prime$}}

\newcommand{\qtl}{\ensuremath{\mathcal{L}}}

\newcommand{\llm}{\ensuremath{\mathcal{M}}}

\definecolor{tug}{RGB}{196,13,30}
\definecolor{perpal}{RGB}{194,9,192}
\newtheorem{example}{Example}

\newcommand{\red}[1]{{\color{tug}#1}}

\newcommand{\textblue}[1]{\texttt{\textcolor{blue}{#1}}}

\newcommand{\textbperpal}[1]{\texttt{\textcolor{perpal}{#1}}}

\newcommand\vldbdoi{10.14778/3828612.3828639}
\newcommand\vldbpages{2880 - 2893}
\newcommand\vldbvolume{19}
\newcommand\vldbissue{10}
\newcommand\vldbyear{2026}
\newcommand\vldbauthors{\authors}
\newcommand\vldbtitle{\shorttitle} 
\newcommand\vldbavailabilityurl{https://github.com/CoDS-GCS/ReSequel}
\newcommand\vldbpagestyle{empty}

\usepackage{afterpage}
\newlength{\oldtextfloatsep}

\usepackage{flushend}
\usepackage[a-2b]{pdfx}

%% file: 0_Abstract.tex
\begin{abstract}
% 1. State the problem
Heuristic query rewriting has long complemented cost-based optimization to improve performance. Such rewrites transform SQL queries into semantically equivalent forms that are easier or faster to execute. Examples are standardizing expressions, eliminating redundancy, propagating constants, pushing down selections and projections, unnesting queries, and utilizing constraints. 
% 2. Say why it's an interesting problem
Modern DBMSs implement hundreds to thousands of such rules, but maintaining them is notoriously difficult. The interactions among rules are complex, and their static nature and application order prevent adaptation to specific query and database characteristics. Recent approaches that use large language models (LLMs) for query rewriting show promise but face challenges regarding the large search space, reliable query verification, and exploitation of metadata.  
% 3. Say what your solution achieves
We present \textbf{ReSequel}, an \emph{outer optimization layer} on top of existing DBMSs to rewrite SQL queries using LLMs. ReSequel leverages catalog and statistical metadata to infer \emph{template-specific rules} that guide the LLM toward effective query transformations. We generate, verify, and rank rewritten query variants on sampled data to ensure result correctness and runtime improvements. 
% 4. Say what follows from your solution
Our experiments cover eight benchmarks: JOB, TPC-H, Stats(-CEB), Public BI, IMDB, DSB, and StackOverflow; multiple DBMSs: PostgreSQL, MySQL, and DuckDB; as well as LLM-based query rewriting baselines. 
ReSequel yields workload-level speedups of up to 16x over native DBMSs and 22x over LLM-based systems, with individual queries exceeding 600x, across eight benchmarks and three DBMSs.
\looseness=-1
\end{abstract}

%% file: setting_paper.tex
\pagestyle{\vldbpagestyle}
\begingroup\small\noindent\raggedright\textbf{PVLDB Reference Format:}\\
\vldbauthors. \vldbtitle. PVLDB, \vldbvolume(\vldbissue): \vldbpages, \vldbyear.\\
\href{https://doi.org/\vldbdoi}{doi:\vldbdoi}
\endgroup
\begingroup
\renewcommand\thefootnote{}\footnote{\noindent
This work is licensed under the Creative Commons BY-NC-ND 4.0 International License. Visit \url{https://creativecommons.org/licenses/by-nc-nd/4.0/} to view a copy of this license. For any use beyond those covered by this license, obtain permission by emailing \href{mailto:info@vldb.org}{info@vldb.org}. Copyright is held by the owner/author(s). Publication rights licensed to the VLDB Endowment. \\
\raggedright Proceedings of the VLDB Endowment, Vol. \vldbvolume, No. \vldbissue\ %
ISSN 2150-8097. \\
\href{https://doi.org/\vldbdoi}{doi:\vldbdoi} \\
}\addtocounter{footnote}{-1}\endgroup

\ifdefempty{\vldbavailabilityurl}{}{
\vspace{.3cm}
\begingroup\small\noindent\raggedright\textbf{PVLDB Artifact Availability:}\\
The source code, data, and/or other artifacts have been made available at \url{\vldbavailabilityurl}.
\endgroup
}

%% file: 1_Introduction.tex
\section{Introduction}
\label{sec:introduction}

%SQL success --> still sensitive to query formulation
QUEL (in Ingres \cite{HeldS75}, based on tuple calculus) and SEQUEL \cite{ChamberlinB74, ChamberlinAEGLMRW76} (in System~R \cite{AstrahanBCEGGKLMMPTWW76}, based on relational algebra) were competing early relational query languages. Ultimately, SEQUEL first became the de-facto standard and later evolved to the SQL standard.    
The success of SQL and the relational model---despite criticism on specific aspects \cite{BerensonBGMOO95,0001L24}---was largely due to its (1) declarativity (what not how), (2) flexibility (compose arbitrary complex queries), (3) freedom for automatic optimization, and (4) physical data independence. Despite these compelling properties, all DBMS remain stubbornly sensitive to suboptimal query formulations because query optimization is \enquote{a never-ending quest for an increasingly better model and repertoire of optimization and execution techniques} \cite{Winslett03c}, various shortcomings and simplifying assumptions (e.g., the independence assumption and its implication for redundant, correlated predicates) \cite{Lohman14}, and brittle rewrites (e.g., for unnesting sub-queries) \cite{0001K15}.  

\begin{figure}[!t]
    \includegraphics[scale=0.75]{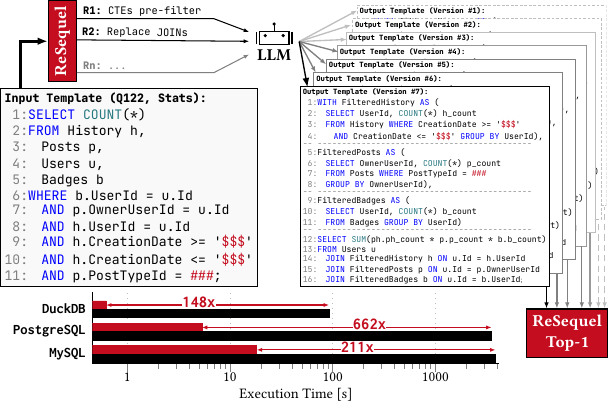}
    \vspace{-0.4cm}
    \caption{\label{fig:example}Comparison of ReSequel's rewriting impact on host DBMS performance. ReSequel rewrites masked templates into multiple variants using LLMs, and then locally verifies them, and selects the $Top\text{-}1$ reconstructed query.}
     \vspace{-0.45cm}
\end{figure}

\textbf{Rule-based Query Rewriting:} Many DBMSs have dedicated rule-based query rewriting systems, which transform queries into equivalent, but more efficient forms. These rule engines date back to the early 1990s \cite{PiraheshHH92,PiraheskLH97}, contain \numprint{1000}s of rules, and are applied heuristically before the cost-based optimization of join orders and physical operators. Example rules are the elimination of \texttt{DISTINCT} if the result contains a \texttt{UNIQUE} attribute, query unnesting, and the propagation of constants over joins (for prefiltering both join sides):
\begin{equation}
R\Join_{R.a=S.b} \left(\sigma_{S.b>7}(S)\right) \text{\small{\ding{212}}} \left(\sigma_{R.a>7}(R)\right)\Join_{R.a=S.b} \left(\sigma_{S.b>7}(S)\right).   
\end{equation}
Such rules are manually crafted by experts over decades~\cite{BegoliCalcite2018} or automatically discovered by specialized tools~\cite{WeTune,QueryBooster}. Since rules are expressed as source-target equivalences of small patterns, whole-query rewriting requires applying rules until a fixpoint or number of iterations. The interaction of rules and their application order renders development and debugging difficult and brittle, requiring considerable time, effort, and expertise. Also, the reliance on predefined rules makes rule engines inherently incomplete, potentially missing valuable rewrite opportunities for new queries~\cite{SlabCity}.

\textbf{LLM-based Query Rewriting:} With the emergence of large language models (LLMs), new approaches to query rewriting appeared, which, however, still face fundamental limitations: 
\begin{enumerate}
\item \emph{LLM-only:} First, there is work on LLM-based query rewriting that submits queries, optional metadata, and instructions to rewrite queries as a whole \cite{LLMSTEER,liu2024query,RBot,DBGpt,tan2025can}. Due to the nature of LLMs, these approaches struggle with hallucinations, and the rewritten queries may underperform due to the lack of database characteristics. 

\item \emph{LLM-based Rule Engines:} LLM-R2~\cite{LLMR2} integrates Apache Calcite~\cite{BegoliCalcite2018} rules, and employs LLMs to refine and select rules for whole query rewriting. This hybrid approach partially addresses the brittleness of rule systems, but the pre-defined rule set cannot exploit unforeseen opportunities.
\end{enumerate}
These categories rely on unverified LLM reasoning or static rule sets, lacking reliability, robust improvements, and flexibility.

\textbf{ReSequel Overview:}  
We introduce ReSequel, a metadata-guided, LLM-assisted query rewriting system that operates as an \emph{outer optimization layer} on top of existing DBMSs. We first gather rich data statistics and metadata on schemas, indexes, constraints, as well as supported DBMS features. ReSequel then generalizes queries into reusable \emph{templates} with masked predicates (for scalable and cost-effective rewriting) and infers operation- and metadata-specific rules. These rules steer the LLM toward promising optimizations relevant for each template such as subquery unnesting, join reordering, or index creation. Each template and its rewriting tasks are encoded into structured prompts that instruct the LLM to generate alternative query variants. ReSequel verifies these variants on sampled data to ensure correctness and ranks them by performance, caching validated rewritten templates. This combination of template generalization, metadata-driven exploration, and sample-based verification enables ReSequel to produce high-performance queries in a robust and scalable manner. Figure~\ref{fig:example} shows an example rewritten template and its performance impact on multiple DBMSs.

\textbf{Contributions:}  
Our primary contribution is the design of ReSequel, a practical framework for holistic query rewriting on top of DBMSs. Our main technical contributions are:\vspace{-0.2cm}
\begin{itemize}
  \item \emph{End-to-end Rewriting Workflow:} We present ReSequel's full pipeline, which integrates catalog metadata, DBMS statistics, and LLM reasoning for query rewriting, as well as verification and ranking of query variants (Section~\ref{sec:architecture}).  

  \item \emph{Template-based Query Generalization:} We introduce query \emph{templates} that mask predicates and cluster similar queries for scalable rewriting. Templates preserve the query structures but enable reuse across queries (Section~\ref{sec:templatization}).  

  \item \emph{Sample-based Verification:} We employ database downsampling and query verification to validate semantic equivalence and select the top-performing query (Section~\ref{sec:downsampling}).  

  \item \emph{Template Rewriting:} We describe the LLM-based template rewriting in terms of constructed prompts, query reconstruction, and query caching. (Section~\ref{sec:rewriting}).

  \item \emph{Experiments:} We study ReSequel's across eight benchmarks 
	and three DBMSs, where we see speedups of up to $3$--$22\times$ over DBMSs and $22\times$ over LLM baselines (Section~\ref{sec:experiments}).  
\end{itemize}

%% file: 2_Background.tex
\section{Background and Challenges}
\label{sec:background}

We review traditional rule-based query rewriting systems and the challenges posed by messy real-world queries. We then formalize the problem of robust query rewriting for LLM-assisted systems.

\begin{figure*}[!t]
    \vspace{-0.65cm}
    \resizebox{.99\textwidth}{!}{\input{figures/fig_Refine_v2}}

    \resizebox{.99\textwidth}{!}{\includegraphics[scale=1]{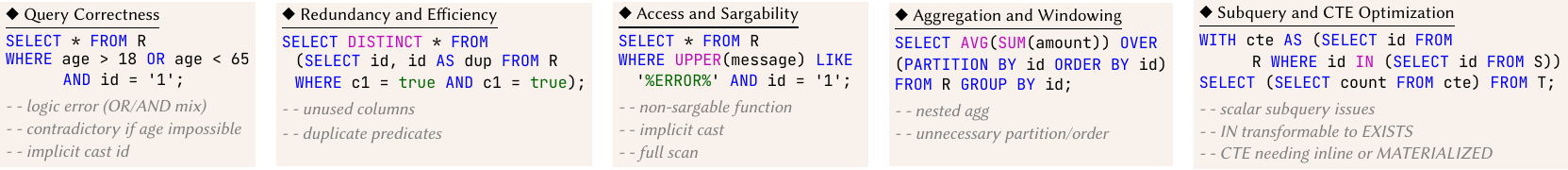}}
    \vspace{-0.35cm}
    \caption{\label{fig:messy_query}Challenges in Phase Ordering and Query Optimization Support in DBMS and Non-DBMS Systems.}
     \vspace{-0.35cm} 
\end{figure*}

\subsection{Rule-based Query Rewriting}

Traditional rule-based query rewriting frameworks~\cite{PiraheshHH92,PiraheskLH97} consist of large collections of equivalence rules (source-to-target patterns) and a rule engine that transforms input queries into simpler or faster forms. Rewriting is typically applied after SQL parsing and semantic analysis but before cost-based query optimization. Rewrites are often implemented as hand-crafted C++ rules, enabling complex transformations such as subquery unnesting~\cite{PiraheskLH97}. Despite their expressiveness, missed optimization opportunities are common.
First, the ordering in which rewrite rules are applied is determined heuristically by experts. As rule sets evolve (e.g., to address specific bugs or performance issues) interactions between rules become difficult to reason about. An incorrect ordering may cause a rewrite to destroy the source pattern required by a more impactful rewrite.
Second, rule sets are applied iteratively until a fixpoint or rewrite budget~\cite{PiraheskLH97} is reached. Under restrictive budgets, large static rule sets often fail to trigger beneficial rewrites. This issue arises because many rules are irrelevant to a given query.
Finally, many rewrite rules operate on few operators and are brittle with respect to source-pattern matching~\cite{Lohman14,0001K15}. Minor variations in query formulation, missing constraints or keys, or differences in operator placement can prevent rewrites from firing~\cite{LohmanOld}. As a result, messy real-world queries remain challenging for rule-based systems.

\subsection{Messy Query Characteristics}

In contrast to many benchmarks, queries in the real world are often messy, originating from SQL-generating applications, text-to-SQL tools, or  inexperienced users. Additionally, database schemas evolve (e.g., splitting structured attributes), yet queries remain unchanged. 

\textbf{Classification:} We characterize common rewrite opportunities in messy queries using a hierarchical classification, shown in Figure~\ref{fig:messy_query}. We distinguish Non-DBMS systems (external rewriting and learned optimizers) and DBMSs by labeling their support as fully supported, unsupported, or conditionally supported (\texttt{+$\textbackslash$-}). Existing solutions address only specific cases, and support remains limited. We also report summary statistics of multiple workloads.

\textbf{Non-DBMS Challenges:} External optimizers primarily improve queries by applying predefined patterns and exploiting metadata, rather than discovering new query formulations. Accordingly, the rewrite effectiveness is highly sensitive to statistics, indexes, and constraints. Despite recent progress, several key challenges remain:
\begin{itemize}
\item \emph{Scalability:} Most rewrite systems operate on individual queries, which does not scale to large workloads. E.g., LearnedRewrite~\cite{LearnedRewrite} requires 16 hours to rewrite the IMDB workload, while workload execution takes only 11 hours. Scalability is even more challenging for LLM-based systems.

\item \emph{Cost:} The monetary cost of LLM-based rewrite systems is often prohibitive, caused by the large number of input/output tokens consumed during query rewriting.

\item \emph{Verification:}
Rewrites may modify queries without guarantees of correctness, but only semantically equivalent queries should be accepted. Moreover, LLM-based systems often apply multiple transformations; for example, by introducing new functions, or extracting subqueries. Existing query verification systems neither support arbitrary transformations nor generalize well across SQL dialects.
\item \emph{Reusability and Robustness:} Most rewrite systems incur substantial upfront overhead, yet their outputs are not reusable, even for highly similar queries. Workload queries are typically instances of few templates, and thus, reusable.
\end{itemize}

\textbf{DBMS Challenges:} DBMSs primarily optimize via cost-based join ordering and operator selection as well as heuristic rewrite systems. However, complex or semantic rewrites often remain the user's responsibility because whole-query rewriting and cost-based search can be expensive. Rewrites may conflict with physical design. For example, applying a function to an indexed column (e.g., \texttt{EXTRACT(YEAR FROM date\_col)}) can prevent index usage unless a matching functional/index expression exists. 

\subsection{Problem Formulation}
To overcome DBMS and Non-DBMS limitations, LLMs offer a promising approach for rewriting messy, text-based queries in a holistic manner. For scalability, recurring query templates are optimized once and reused across query instances. This design enables thorough verification and ranking of query variants on samples.

\textbf{Notation:} Given an input SQL query \Q, our goal is to generate a semantically equivalent, but faster query \QP.
To this end, we define a query rewriting operator
$R = [\mathcal{R}, \mathcal{F}, \mathcal{E}]$,
where $\mathcal{R}$ is a set of predefined semantics-preserving rewrite rules, $\mathcal{F}$ is a set of interface functions which can be overwritten, and $\mathcal{E}$ are exploratory rewrite examples that enable discovering new transformations. The rewriting process is guided by an LLM, which applies $R$ to the input query:
$\QP = R(\Q)$.
All rewritten queries must be semantically equivalent to the original query. To evaluate performance, candidate queries are executed on a downsampled database \ts, and the final output query is selected by minimizing execution latency:
\[
\QP = \arg\min_{\Q_r \equiv Q} \ell(\Q_r, \ts), \ \ \ \ \ \text{s.t., } \ell(\Q_r, \ts) \leq \ell(\Q, \ts).
\]

%% file: figures/fig_Refine_v2.tex
\newcommand{\yes}{\cellcolor{lightgray!70}\fontsize{4pt}{4pt}\selectfont{\color{black}\ding{52}}}
\newcommand{\no}{\cellcolor{lightgray!10}{\color{gray}\ding{55}}}
\newcommand{\yesno}{\cellcolor{lightgray!30}\fontsize{8pt}{8pt}\selectfont{\color{gray!120}+}\tiny\textbf{\textbackslash}\fontsize{8pt}{8pt}\selectfont{\color{gray!120}-}}

\newcommand{\point}{\color{gray}\ding{220}\color{black}}

\begin{tikzpicture}[node distance=1mm, font=\tiny, line width=0.3pt,
    itemstyle/.style={inner ysep=0.4mm, inner xsep=0.3mm,minimum width=65pt, text width=65pt, draw=black},
    subitemstyle/.style={inner ysep=0.4mm, inner xsep=0.3mm,minimum width=115pt, text width=115pt, draw=black}
    ]
     \node [shape=rectangle](table1) {
        %\fontsize{4}{4}\selectfont
        \addtolength{\tabcolsep}{-0.7em}
        \begin{tabular}{ll|c|c|c|c|c|c|c|c|c|c|c|}
        \cline{3-4} \cline{5-5} \cline{7-13} & & \textbf{\texttt{None-DBMS}}   & \textbf{\texttt{DBMS}}& \textbf{\texttt{ReSequel}} & & \textbf{\texttt{TPC-H}} & \textbf{\texttt{JOB}} & \textbf{\texttt{Stats(-CEB)}} & \textbf{\texttt{PBI}} & \textbf{\texttt{StackOverflow}}& \textbf{\texttt{DSB}}& \textbf{\texttt{IMDB}}\\ \cline{3-4} \cline{5-5} \cline{7-13}
        %%%%%%%%%%%
        \multirow{4}{*}{\ \ \ \texttt{\textbf{\makecell{Query\\ Correctness}}}}   & \point\ \texttt{Deterministic results \& logic errors}~\cite{SongDGCZWWWH24,ZhouCPWM021, SatrianiVSRVP25} & \yes   & \no & \yes& & \texttt{0}& \texttt{17.70}& \texttt{9.59}& \texttt{3.00}& \texttt{18.74} &\texttt{11.54}&\texttt{0}\\ \arrayrulecolor{lightgray!60}\cline{7-13}
                                    & \point\ \texttt{Ambiguous joins, non-deterministic sorts}~\cite{ZhaoRDW16,Meimarakis21}              & \yesno & \no & \yes& & \texttt{22.73}& \texttt{28.32}& \textbf{\red{\texttt{92.47}}}& \texttt{2.00}& \texttt{20.50} &\texttt{42.31}&\texttt{37.95}\\  \arrayrulecolor{lightgray!60}\cline{7-13}
                                    & \point\ \texttt{Contradictory/always-true predicates}\cite{LivshitsKTIKR21,MullerWL23}                  & \no    & \no & \yes & &\texttt{0}& \texttt{10.62}& \texttt{61.64}& \texttt{12.00}& \texttt{0} &\texttt{7.69}&\texttt{5.42}\\  \arrayrulecolor{lightgray!60}\cline{7-13}
                                    & \point\ \texttt{Implicit casts}~\cite{FurstKNZS25,pgdoc_create_cast}& \no  & \no & \yes& &\texttt{0}& \texttt{14.16}& \texttt{0}& \texttt{1.00}& \texttt{0} &\texttt{1.92}&\texttt{0}\\  \arrayrulecolor{lightgray!60}\cline{7-13}
                                    & & & & &&&&&&&&\\  \arrayrulecolor{lightgray!60}\cline{7-13}
        %%%%%%%%%%%%                            
        \multirow{3}{*}{\ \ \ \texttt{\textbf{\makecell{Redundancy and \\ Efficiency}}}} & \point\ \texttt{Dead code \& redundant operations}~\cite{FoufoulasS23,SchmidtF022,MarklRSLP04,BaumstarkJS23}    & \no   & \yesno & \yes & & \texttt{31.82}& \texttt{15.93}& \texttt{21.92}& \textbf{\red{\texttt{42.00}}}& \texttt{10.17}&\textbf{\red{\texttt{51.92}}}&\texttt{3.39}\\  \arrayrulecolor{lightgray!60}\cline{7-13}
                                    & \point\ \texttt{Unused columns, and duplicate predicates/joins}~\cite{DreselerBRU20,MullerWL23,FreitagBSKN20}  & \yesno & \yesno & \yes & & \texttt{13.64}& \texttt{100.00}& \texttt{3.42}& \texttt{18.00}& \texttt{0.25}&\texttt{26.92}&\textbf{\red{\texttt{72.85}}}\\  \arrayrulecolor{lightgray!60}\cline{7-13}
                                    & \point\ \texttt{Simplify DISTINCT clauses and unnecessary nesting}~\cite{HaffnerD23,BakkeK16}    & \yesno & \yesno & \yes& & \texttt{13.64}& \texttt{0}& \texttt{0}& \texttt{33.00}& \texttt{0.34}&\texttt{3.85}&\texttt{0}\\  \arrayrulecolor{lightgray!60}\cline{7-13}
                                    & & & & &&&&&&&&\\  \arrayrulecolor{lightgray!60}\cline{7-13}
        %%%%%%%%%%%%                            
        \multirow{6}{*}{\ \ \ \texttt{\textbf{\makecell{Access and\\ Sargability}}}}       & \point\ \texttt{Predicates for optimal index usage}~\cite{KesterAI17,DittrichNS21,HershcovitchKW022}   & \yesno & \yesno & \yesno& & \texttt{4.55}& \texttt{17.70}& \texttt{0}& \texttt{15.00}& \textbf{\red{\texttt{20.84}}}&\texttt{3.85}&\texttt{27.60}\\  \arrayrulecolor{lightgray!60}\cline{7-13}
                                    & \point\ \texttt{Functions-on-columns to range predicates}~\cite{BrunoC06,Zhou0Z0W24,VogelsgesangML019}  & \yesno & \no  & \yesno& & \texttt{4.55}& \texttt{3.54}& \texttt{0}& \texttt{33.00}& \texttt{0}&\texttt{26.92}&\texttt{27.68}\\  \arrayrulecolor{lightgray!60}\cline{7-13}
                                    & \point\ \texttt{Recommend expression/partial indexes}~\cite{HeerenJP03,SiddiquiW23,WangLLBLW24,ChakkappenKMKSZZLZ25}      & \yesno & \yesno & \yesno & & \texttt{9.09}& \texttt{0}& \texttt{0.68}& \texttt{19.00}& \texttt{11.60}&\texttt{21.15}&\texttt{5.16}\\  \arrayrulecolor{lightgray!60}\cline{7-13}
                                    & \point\ \texttt{Replace LIKE '\%x' with trigram/full-text patterns}~\cite{QueryBooster,DittrichN20,GretscherD25}    & \yesno & \no & \yesno& & \texttt{18.18}& \textbf{\red{\texttt{64.60}}}& \texttt{0}& \texttt{0}& \texttt{0}&\texttt{0}&\texttt{28.89}\\  \arrayrulecolor{lightgray!60}\cline{7-13}
                                    & \point\ \texttt{Normalize types to avoid casts}~\cite{HuangSL023,OusterhoutMDFRT23}    & \no  & \no & \yes& & \texttt{0}& \texttt{11.50}& \texttt{7.53}& \texttt{25.00}& \texttt{0}&\texttt{5.77}&\texttt{4.29}\\  \arrayrulecolor{lightgray!60}\cline{7-13}
                                    & \point\ \texttt{Trim SELECT * to needed columns}~\cite{DreselerBRU20,FreitagBSKN20}   & \no & \no & \yes & & \texttt{9.09}& \texttt{0}& \texttt{0}& \texttt{0}& \texttt{0}&\texttt{1.92}&\texttt{0}\\  \arrayrulecolor{lightgray!60}\cline{7-13}
                                    & & & & &&&&&&&&\\  \arrayrulecolor{lightgray!60}\cline{7-13}
         \multirow{2}{*}{\ \ \ \texttt{\textbf{Join Logic}}}              &\point\ \texttt{Heuristic join reordering}~\cite{YanUL23,LeisRGMBKN18}  & \yesno & \yesno & \yesno& & \texttt{4.55}& \texttt{19.47}& \texttt{19.18}& \texttt{0}& \texttt{0.42}&\texttt{23.08}&\texttt{67.98}\\  \arrayrulecolor{lightgray!60}\cline{7-13}
                                    &\point\ \texttt{Change join types}~\cite{TziavelisGR20,FreitagBSKN20} & \no & \no & \yes & &\texttt{4.55}& \texttt{0}& \texttt{9.59}& \texttt{0}& \texttt{0}&\texttt{5.77}&\texttt{0}\\ \arrayrulecolor{lightgray!60}\cline{7-13}
                                    & & & & &&&&&&&&\\  \arrayrulecolor{lightgray!60}\cline{7-13}
        %%%%%%%%
        \multirow{3}{*}{\ \ \ \texttt{\textbf{\makecell{Aggregation and\ \ \ \\ Windowing}}}} &\point\ \texttt{Collapse nested aggregates/windows}~\cite{FentBN23,ChasialisFSI26}  & \no & \no & \yes & & \texttt{4.55}& \texttt{0}& \texttt{0}& \texttt{0}& \texttt{0}&\texttt{5.77}&\texttt{0}\\  \arrayrulecolor{lightgray!60}\cline{7-13}
                                    &\point\ \texttt{Replace DISTINCT-on-join}~\cite{BirlerKN24,JustenRFLTLBHZMB24}  & \no & \no & \yes & & \texttt{4.55}& \texttt{0.88}& \texttt{1.37}& \texttt{0}& \texttt{2.44}&\texttt{1.92}&\texttt{0}\\  \arrayrulecolor{lightgray!60}\cline{7-13}
                                    &\point\ \texttt{Remove unnecessary PARTITION/ORDER}~\cite{LeisKK015,KlabeS23}  & \no & \no & \yes& &\texttt{4.55}& \texttt{2.66}& \texttt{0}& \texttt{0}& \texttt{0}&\texttt{0}&\texttt{0}\\  \arrayrulecolor{lightgray!60}\cline{7-13}
                                    & & & & &&&&&&&&\\  \arrayrulecolor{lightgray!60}\cline{7-13}
        %%%%%%%%
        \multirow{3}{*}{\ \ \ \texttt{\textbf{\makecell{Subquery and \\ CTE Optimization}}}} &\point\ \texttt{Scalar subqueries}~\cite{abs-2406-06886,KossmannPN23}  & \yes & \yesno & \yes&& \textbf{\red{\texttt{27.27}}}& \texttt{0}& \texttt{0}& \texttt{0}& \texttt{1.76}&\texttt{17.31}&\texttt{0}\\  \arrayrulecolor{lightgray!60}\cline{7-13}
                                    &\point\ \texttt{Transform IN/NOT IN to EXISTS/NOT EXISTS}~\cite{ShankhdharLNSSA24,FejzaGL24}  &\yes &\yes & \yes& & \texttt{18.18}& \texttt{0}& \texttt{0.68}& \texttt{0}& \texttt{0}&\texttt{9.62}&\texttt{1.75}\\  \arrayrulecolor{lightgray!60}\cline{7-13}
                                    &\point\ \texttt{Inline CTEs or mark MATERIALIZED}~\cite{SichertN22,FloratosGSCZ21}  & \yes & \yesno & \yes& & \texttt{4.55}& \texttt{0}& \texttt{0}& \texttt{0}& \texttt{0}&\texttt{19.23}&\texttt{0}\\ \arrayrulecolor{black} \cline{3-4}\cline{5-5}  \cline{7-13}\cline{7-13}

        \end{tabular}   
    };    

    \node[left=of table1, inner sep=1mm, draw=black, xshift=-5pt, yshift=43pt, rotate=90, fill=lightgray!10] (qq){        
            \textbf{\texttt{Messy Query Rewrite Objectives}}
    };

    \node[left=of table1, inner sep=1mm, draw=none, xshift=0pt, yshift=4.3pt, fill=none] (q){};

    \draw[gray,{Triangle[width=2*2pt,length=3pt]}-{Triangle[width=2*2pt,length=3pt]}, line width=0.7pt]  ($(q.north)+(14pt, 56.5pt)$) -- ($(q.north)+(9pt, 56.5pt)$) -- ($(q.north)+(9pt, -78.5pt)$) -- ($(q.north)+(14pt, -78.5pt)$);
    \draw[gray,{Triangle[width=2*2pt,length=3pt]}-, line width=0.7pt]  ($(q.north)+(14pt, 29.5pt)$) -- ($(q.north)+(9pt, 29.5pt)$);
    \draw[gray,{Triangle[width=2*2pt,length=3pt]}-, line width=0.7pt]  ($(q.north)+(14pt, -3.5pt)$) -- ($(q.north)+(3pt, -3.5pt)$);
    \draw[gray,{Triangle[width=2*2pt,length=3pt]}-, line width=0.7pt]  ($(q.north)+(14pt, -33.5pt)$) -- ($(q.north)+(9pt, -33.5pt)$);
    \draw[gray,{Triangle[width=2*2pt,length=3pt]}-, line width=0.7pt]  ($(q.north)+(14pt, -54.5pt)$) -- ($(q.north)+(9pt, -54.5pt)$);
    % \draw[gray, line width=0.7pt]  ($(qq.south)+(0pt, -3.5pt)$) -- ($(qq.south)+(16pt, -3.5pt)$);
    %%%%%%%%%%%%%%%%%%%%%%%%%%%%%%%%%%%%%%%%%%%%%%%%%%%%%%%%%%%%%%%%%%
    \draw[gray,-{Bar[width=18.4pt, line width=1pt]}, line width=0.5pt]  ($(q.north)+(47pt, 56.5pt)$) -- ($(q.north)+(62pt, 56.5pt)$);
    \draw[gray,-{Bar[width=12.4pt, line width=1pt]}, line width=0.5pt]  ($(q.north)+(53pt, 29.4pt)$) -- ($(q.north)+(62pt, 29.4pt)$);

    \draw[gray,-{Bar[width=30pt, line width=1pt]}, line width=0.5pt]  ($(q.north)+(44pt, -3.5pt)$) -- ($(q.north)+(62pt, -3.5pt)$);
    \draw[gray,-{Bar[width=6.2pt, line width=1pt]}, line width=0.5pt]  ($(q.north)+(43pt, -33.5pt)$) -- ($(q.north)+(62pt, -33.5pt)$);

    \draw[gray,-{Bar[width=12.8pt, line width=1pt]}, line width=0.5pt]  ($(q.north)+(57pt, -54.7pt)$) -- ($(q.north)+(62pt, -54.7pt)$);
    \draw[gray,-{Bar[width=12.4pt, line width=1pt]}, line width=0.5pt]  ($(q.north)+(57pt, -78.6pt)$) -- ($(q.north)+(62pt, -78.6pt)$);

\end{tikzpicture}

%% file: 3_Architecture.tex
\vspace{-0.4cm} %somehow needed, even though enough space
\section{ReSequel System Overview}
\label{sec:architecture}

\begin{figure}[!t]
    \resizebox{.99\columnwidth}{!}{\includegraphics[scale=1]{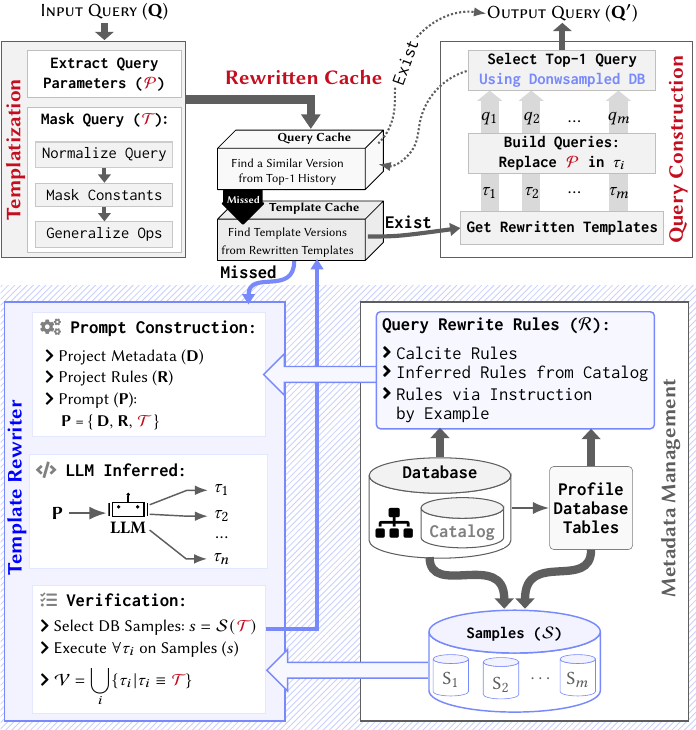}}
    \vspace{-0.35cm}
    \caption{\label{fig:resequel_arch}ReSequel Architecture and Workflow.}
    \vspace{-0.45cm}
\end{figure}

ReSequel is an end-to-end SQL query rewriting system designed as an outer optimization layer for existing DBMSs. It employs zero-shot, in-context learning to guide LLMs in generating semantically equivalent queries with improved performance. ReSequel relies on explicit task specification as well as operation- and metadata-aware guidance derived from system metadata, data characteristics, and DBMS-specific features. As illustrated in Figure~\ref{fig:resequel_arch}, ReSequel consists of modular components that support query templatization, metadata management, guided rewriting, and scalable construction and evaluation of executable query variants.
Rather than optimizing individual ad-hoc queries, ReSequel performs rewriting at the level of workload templates. To support this workflow, the system distinguishes between two layers: \emph{online query processing} and \emph{offline query rewriting}. Queries whose templates have been processed are handled by the online layer, which applies cached rewrites with minimal overhead. Queries of unseen templates are routed to the offline layer, where ReSequel performs metadata-driven rewrite generation, verification, and template materialization, enabling subsequent queries to benefit from reusable, optimized templates.

\vspace{-0.1cm}
\subsection{Query, Metadata, and Rewrite Hints} 
\begin{figure}[!t]
    \resizebox{.99\columnwidth}{!}{\includegraphics[scale=1]{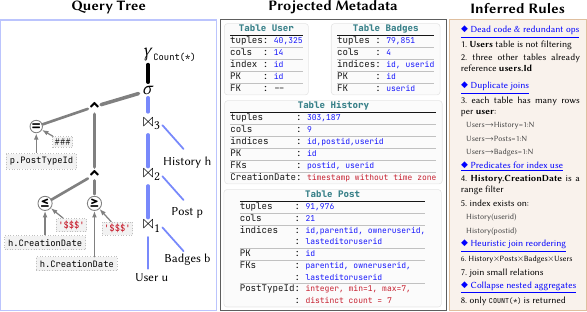}}
    \vspace{-0.35cm}
    \caption{\label{fig:example_q122_metadata} Materialization Steps for STATS $Q\#122$ (Figure~\ref{fig:example}).}
     \vspace{-0.45cm}
\end{figure}

For an input query, we first collect metadata related to its parameters and associate it with categories of messy query characteristics. We extract the query template and model each query as a query tree composed of specific operators (e.g., Join, Filter, Projection). Based on the referenced tables, we prepare the metadata, including attribute characteristics, indexes, and primary and foreign keys. Finally, we check a set of predefined conditions to identify and label the query with specific characteristics, which can be used as rewriting hints. Figure~\ref{fig:example_q122_metadata} shows an example of these components.

\subsection{Online Query Processing}
In the online layer, each incoming SQL query is templatized and then matched against cached templates to identify reusable optimized variants. Upon a cache hit, ReSequel reconstructs the query and returns an optimized version \QP. Key online components are:

\textbf{Templatization:} Query workloads in production systems are rarely ad-hoc. They typically arise from a small number of parameterized SQL templates instantiated with different constants. For example, the IMDB workload contains \numprint{13646} queries but only 79 distinct templates.  
ReSequel therefore rewrites templates instead of individual queries to achieve scalability and low overhead. It canonicalizes each incoming query by masking literal constants and normalizing syntactic variants. Queries with identical logical structure are mapped to a common template identifier that indexes cached templates. Templatization also extracts query parameters and aligns them with placeholders, enabling later reconstruction of query instances.  
ReSequel is designed to operate under high database load. Template-based rewriting is applied upfront while remaining flexible enough to handle previously unseen queries.  

\textbf{Caching:}
ReSequel maintains two caches to support efficient reuse. The \texttt{Query Cache} stores rewritten instances of previously optimized queries; upon a cache hit, ReSequel substitutes query parameters and immediately returns the optimized query \QP. If no match is found, the query is matched against the \texttt{Template Cache}, which stores verified rewritten variants per template. On a template hit, ReSequel invokes query construction. Queries without cached templates are forwarded to the offline rewriting layer. Reusing optimized templates across workloads as well as the discovery of new built-in rewrites is interesting future work.

\textbf{Query Construction:}
Given a set of rewritten and verified template variants ($\mathbf{t} = \{\tau_1, \tau_2, \ldots, \tau_m\}$),
ReSequel reconstructs query instances by substituting the extracted parameters of the original query \Q\ into the corresponding placeholders of each template variant. These candidate query instances ($\mathbf{q} = \{q_1, q_2, \ldots, q_m\}$) are subsequently evaluated and ranked on downsampled data, and we select the top-1 candidate as the rewritten query.

\subsection{Offline Query Rewriting}
Previously unseen query templates undergo offline rewriting and verification. This process can be executed once upfront for a workload, asynchronously off the critical path, or on demand.

\begin{figure*}[!t]
    \vspace{-0.65cm}
    \centering     
    \subfigure[Q1: A Query with a Generalized Operator]{\resizebox{.45\textwidth}{!}{\includegraphics[scale=1]{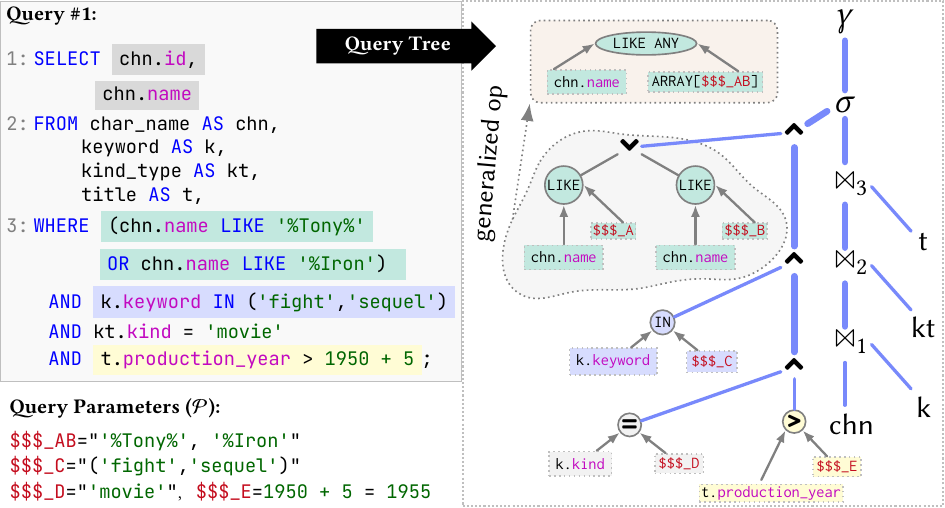}}}  
    \hfill 
    \subfigure[Q2: A Query with a Generalized Operator and Redundant Filter]{\resizebox{.52\textwidth}{!}{\includegraphics[scale=1]{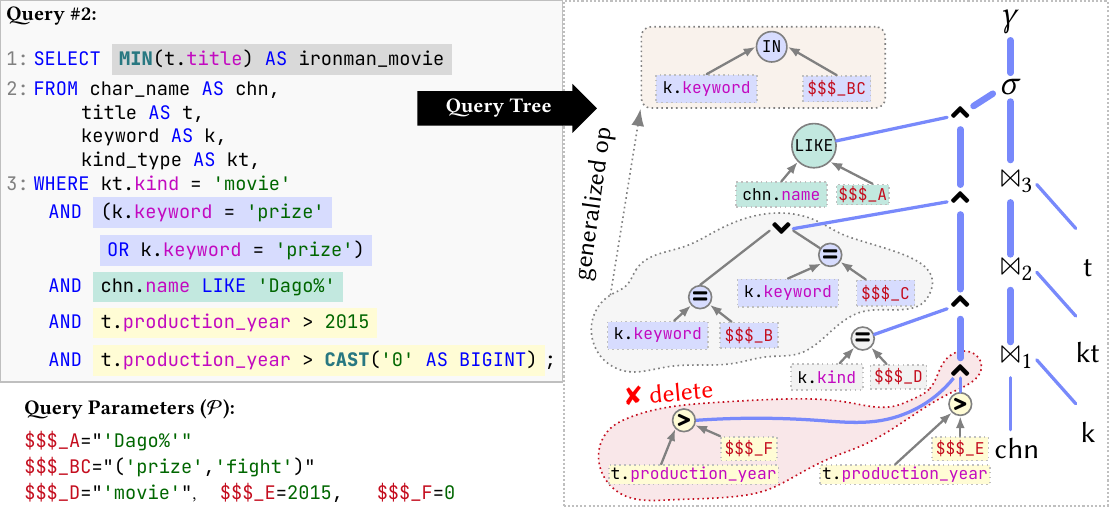}}}
    \vspace{-0.55cm}  
    \caption{Example of Normalization, Masking, and Generalization of Operators for Two Queries / Parameter Materialization.}
    \label{fig:template-example-q1-2}
    \vspace{-0.2cm}    
\end{figure*}

\textbf{Metadata Management:}
The metadata management component creates a unified representation of metadata, rewrite rules, and verification samples required for template rewriting. This information is projected into LLM prompts to guide rewrite generation.

\begin{itemize}
\item \emph{DBMS Catalog and Data Profiling.}
ReSequel leverages DBMS catalog metadata, including schemas, constraints, and statistics, together with additional table-level profiling to support rewrite decisions that depend on data characteristics.

\item \emph{Query Rewrite Rules ($\mathbf{R}$).}
ReSequel incorporates static (heuristically applied) and cost-based (dependent on statistics) rewrite rules. The system integrates rules from query optimizers (e.g., Calcite~\cite{Calcite}), profiled data, and curated, example-based guidance into a unified rule representation. We have collected nearly 50 examples of meta rules, crafted by hand to guide the LLM in query rewriting and exploratory search.

\item \emph{Database Downsampling ($\mathcal{S}$).}
To enable fast verification of rewritten queries, ReSequel constructs representative downsampled table clusters for each template. %The design and rationale of the downsampling process are detailed in Section~\ref{sec:downsampling}.
\end{itemize}

\textbf{Template Rewriter:}
ReSequel performs template rewriting to amortize rewrite costs across many query instances. It does not assume that a single rewrite is optimal for all template instantiations. Instead, the system selects the best variant per query instance based on actual predicate values during Top-1 selection.
Given a template $\qt$ and its associated metadata, ReSequel formulates a rewrite task and invokes the LLM. The prompt
($\mathbf{P} = \{\mathbf{D}, \mathbf{R}, \qt\}$)
combines projected metadata, rewrite rules, and the template text. The LLM generates template variants, which are verified on downsampled data. Valid variants are then stored in the template cache for reuse.

\textbf{LLM Dependency:} ReSequel collects metadata and rules, and generalizes workload queries into templates to reduce the reliance on LLMs. We do not directly trust LLM outputs. Instead, the LLM is used to create rewritten variants, but deterministic components validate them. Therefore, model evolution mainly affects the diversity of rewritten variants, not their ranking or correctness.

%% file: 4_Templatization.tex
%\setspace{6pt} % reduce space after algorithm

\section{Workload Templatization}
\label{sec:templatization}

ReSequel's query rewriting is based on the templatization of queries for scalable and cost-effective rewriting. Figure~\ref{fig:template-example-q1-2} shows two example queries that are mapped to a single template. Templatization consists of two steps: template extraction and template refinement.
%%
% to address SIGMOD comments related to:
%templatization may lose rewrite opportunities or correctness
%
ReSequel’s templatization preserves the structural semantics of queries, including join graphs, grouping and aggregation operators, and predicate structure (columns and operators). Only literal constants and variable-length predicate lists are generalized. As a result, templatization defines the scope for rewriting.

\textbf{Template Extraction:} First, we extract the predicate values of individual filtered attributes and stored them in \p. Additionally, we refine the values by removing duplicate filter values, eliminating unnecessary conditions (e.g., \texttt{CAST('0' AS BIGINT) = 0}), and performing constant folding (i.e., executing and replacing simple arithmetic operations). This refinement is necessary because most database systems struggle with overly complex predicates. Subsequently, we generalize and mask the templates to create an abstract view \qt and thus, reduce the number of distinct templates. Algorithm~\ref{alg:templatization} shows the pseudo-code for the two templatization steps. We first extract all predicate values from the queries in Line~\ref{get:queryparam}. ReSequel identifies a key pattern for each actual value (Line~\ref{key_embeding}). This key pattern is a unique string obtained from the SQL query to extract the actual value or to find and replace it during query reconstruction. After finding unique keys for all values, we identify the operation type (e.g., \texttt{IN}, \texttt{LIKE}) between the value and referenced attribute (see Line~\ref{find_op}), and materialize these key patterns in Line~\ref{save_params}. ReSequel refines workload templates to produce a compact set of distinct templates. Structural constructs---such as outer joins with NULL-sensitive semantics, correlated subqueries, and window functions---are retained unmodified, allowing the LLM to apply rewriting rules over the full context. Order-sensitive and -insensitive queries are merged into a single template during templatization and separated during query reconstruction and Top-1 selection. We are using SQLGlot~\cite{sqlglot} to parse SQL queries into an AST and traverse it to identify operators to merge and extract predicate values. Therefore, dialect support, as well as support for specific constructs, depends on the capabilities provided by SQLGlot. A limitation lies in dialect-specific operator coverage: unrecognized constructs cannot be generalized into semantically equivalent templates. Broad and extensible support of SQL dialects is an
interesting direction for future work.\shorten

\begin{algorithm}[!t] \fontsize{7.9}{0}
\caption{\textsc{templatization}$(\Q)$}\label{alg:templatization}
  
\begin{algorithmic}[1]
  \REQUIRE{SQL Query \Q}
  \ENSURE{Query Template \qt, Query Parameters \p}
    \STATE \p $\leftarrow \text{dict}();\quad \quad  \qt \leftarrow \varnothing;$ \label{def:param}
		\COMMENTLINE{Phase 1: Template Extraction}
    \STATE V $\leftarrow \textsc{extractQueryValues}(\Q)$\label{get:queryparam} \algorithmiccomment{get query parameter values.}
    \STATE V$ ^{\prime}\leftarrow \textsc{refineQueryValues}(\text{V})$\label{refinequeryparam} \algorithmiccomment{e.g., remove duplicate values.}
    \FOR[iterate over query values.]{$v \in \text{V}^{\prime}$}{ \label{itr:queryvalues}
        \STATE key $\leftarrow \textsc{getKeyPattern}(\Q,v)$ \label{key_embeding}\algorithmiccomment{key embedding.}
        \STATE op $\leftarrow \textsc{getOperation}(\Q,\text{key})$\label{find_op}\algorithmiccomment{operation identification.}
        \STATE \p $[\text{key}]\leftarrow \{\text{op}, v\}$\label{save_params}\algorithmiccomment{add parameter value to \p.}
     }\ENDFOR
     \COMMENTLINE{Phase 2: Template Refinement}
     \STATE t $\leftarrow \textsc{normalizeQuery}(\Q)$\label{normalizequery} \algorithmiccomment{query normalization.}
     \STATE t $\leftarrow \textsc{maskQueryValues}(\text{t})$\label{maskquery} \algorithmiccomment{e.g., mask all strings to '\$\$\$'.}
     \STATE \qt $\leftarrow \textsc{generalizeQueryOperations}(\text{t},\p)$\label{querygeneralization}
    \RETURN  $\qt, \p$
\end{algorithmic}
\end{algorithm}
%\resetspace % reset space after algorithmc 

\textbf{Template Refinement:} Second, ReSequel extracts query templates through a three-step refinement process. In Step 2a (Line~\ref{normalizequery}), we normalize the original queries by sorting the query outputs (since the order of projected columns is irrelevant), reordering join operations using a fixed alphabetical pattern for uniformity, and replacing table aliases with their original, non-conflicting names. To avoid complex variable handling, we extract a query tree from each query and derive the template from this tree.
In step 2b in Line~\ref{maskquery}, ReSequel replaces all actual values with masked values. For example, we use the pattern \texttt{$\$\$\$\_A$} for strings and \texttt{$\#\#\#\_A$} for numerical values. After this step, we obtain a nearly static query template.
Additionally, in step 2c in Line~\ref{querygeneralization}, we generalize the template's operations. Applications may inject flexible filter values (e.g., \texttt{LIKE ANY (ARRAY ['val1', 'val2', ...]}). If we keep the original operations, each different size of value lists would result in a distinct template. To generalize such lists, we replace them with a single masked value (e.g., \texttt{LIKE ANY (ARRAY [$\$\$\$$])}). Similarly, for fixed operations such as \texttt{col1 = 1 OR col1 = 2 OR col1 = 3}, we generalize them into a list operation of \p\ values and replace them with a single masked value.
This refinement substantially reduces the number of query templates to rewrite.
Templatization does not reason about rewriting query instance; it only ensures scalable and reusable rewriting, while later, we select optimal variants.

\begin{figure}[!t]
\vspace{-0.25cm} 
    \centering  
    \resizebox{.99\columnwidth}{!}{\includegraphics[scale=1]{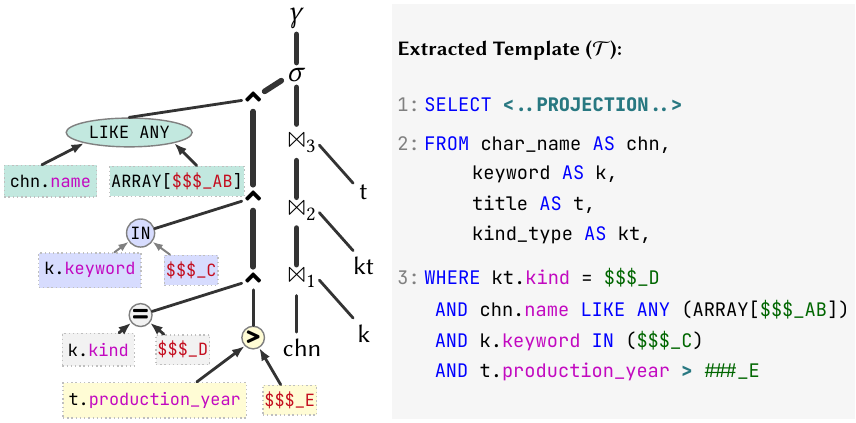}}   
    \vspace{-0.25cm}  
    \caption{Extracted Template from Queries \#1 $\&$ \#2 in Figure~\ref{fig:template-example-q1-2}.}
    \label{fig:template-example}
    \vspace{-0.45cm}    
\end{figure}

\begin{example}[Workload Templatization] 
Figure~\ref{fig:template-example} illustrates an example of extracting a template for the two SQL queries (in PostgreSQL dialect) from Figure~\ref{fig:template-example-q1-2}. Here, distinct parts are highlighted in different colors. 
%
% First, the queries contain different projection clauses, which we mask during template extraction. We assume that the \texttt{SELECT} clause (i.e., projected columns and aggregation) has little impact on runtime. However, the original projection is restored during reconstruction. 
%Essam
% a reviwer may think it is incorrect performance claims
%instead let's say
First, the queries contain different projection clauses, which are masked during template extraction to enable structural clustering. The original projection (including projected columns and aggregation) is restored during query reconstruction and top-1 selection.
Second, the order of clauses does not matter; for example, \texttt{kt.kind} appears in different positions across queries. 
Third, ReSequel generalizes the \texttt{LIKE}, \texttt{OR}, and \texttt{IN} operators. At this stage, we do not attempt to determine whether \texttt{k.keyword IN ('fight','sequel')} performs worse than \texttt{k.keyword = 'fight' OR k.keyword = 'sequel'}. Our goal is simply to generalize such operations.
%Essam
This generalization preserves predicate semantics while avoiding template explosion due to syntactic variations.
Fourth, the extracted query parameters reveal that the filter condition \texttt{t.production\_year > CAST('0' AS BIGINT)} in Query \#2 is redundant and can be removed.
Fifth, ReSequel's template extraction and refinement enables merging queries into a single template when the same column appears with different operations. 
%
% In Query \#1 and Query \#2 (after removing unnecessary conditions), ReSequel performs outer generalization to merge \texttt{LIKE} and \texttt{LIKE ANY}, creating a unified leaf condition in the query tree. Finally, the query tree is converted into an SQL query to produce the template, and we materialize the key pattern and operation for each actual value in a query parameter table for later reconstruction.
In Query~\#1 and Query~\#2, after removing unnecessary conditions, ReSequel applies outer generalization to merge \texttt{LIKE} and \texttt{LIKE ANY} into a unified leaf condition. The resulting query tree is then converted into an SQL template, and key patterns and operations are materialized for later query reconstruction.
\end{example}

%% file: 5_Downsampling.tex
\section{Database Downsampling}
\label{sec:downsampling}
With the workload templates in hand, workload verification and costing requires executing actual queries over samples. ReSequel extracts multiple datasets with different schema subsets and stores them separately. We use downsampled datasets for three reasons: (1) to perform syntax checks of LLM-generated queries, (2) to verify LLM-rewritten queries for result correctness, and (3) to evaluate the runtime of verified candidate queries, and select the top-1 candidate per original query. Drawing multiple samples per cluster of tables ensures robust verification and performance evaluation. Algorithm~\ref{alg:downsampling} shows the overall downsampling process.

\textbf{Downsampling Scope and Guarantees:}
ReSequel uses downsampled databases to filter invalid rewritten queries and to evaluate performance among candidate queries. Downsampling does not provide a formal guarantee of semantic equivalence or exact runtime behavior on the full database. To mitigate this limitation, ReSequel verifies candidates across multiple samples per schema cluster, and includes the original query as a fallback.
For analyzing the downsampling trustworthiness, we conduct dedicated tests. We execute the queries on sample datasets and collect the results. Next, we randomly remove some query results from selected samples and re-executed the queries to verify that the outputs of rewritten and original queries remain identical. Here, we preserve predicate matches to avoid empty query results equivalence checking.

%\setspace{6pt} % reduce space after algorithm
\begin{algorithm}[!t] \fontsize{7.9}{0}
\caption{\textsc{databaseSampling}$(\qtl, \D, \cat, \tau)$}\label{alg:downsampling}
  
\begin{algorithmic}[1]
  \REQUIRE{List of Templates \qtl[$\qt_1,...,\qt_n$], Database \D, Data Catalog \cat,\\ Database Instance Limits $\tau$}
  \ENSURE{Set of Databases \ts}
    \STATE X $\leftarrow$ dict() \algorithmiccomment{template info dictionary.}
    \COMMENTLINE{Phase 1: Schema Selection}
		\FOR[iterate over workload templates.]{$t \in \text{\qtl}$}{ \label{itr:templates}
        \STATE tbls $\leftarrow \textsc{getTables}(t)$ \label{get_tables}\algorithmiccomment{get all table names in template.}
        \STATE cols $\leftarrow \textsc{getColumnNames}(t)$\label{get_cols}\algorithmiccomment{get all col names in template.}
        \STATE schema $\leftarrow \textsc{getSchema}(\D,\cat,\text{tbls})$\label{get_templateschema}\algorithmiccomment{get template schemas.}
        \STATE X$[t]\leftarrow$ (tbls, cols, schema) \label{save_info}
     }\ENDFOR
     \COMMENTLINE{Phase 2: Schema Clustering}
     \STATE C $\leftarrow \textsc{schemaClustering}(\text{X})$ \label{templateclustering} \algorithmiccomment{cluster templates.}
     \COMMENTLINE{Phase 3: Database Sampling}
     \STATE \ts $\leftarrow$ dict()  
     \FOR[iterate over template clustrs.]{$c \in \text{C}$}{ \label{itr:cluster}
        \STATE tbls $\leftarrow \textsc{getTables}(c)$ \label{get_clustertables}\algorithmiccomment{get table names in cluster.}
        \STATE cols $\leftarrow \textsc{getColumnNames}(c)$\label{get_clustercols}\algorithmiccomment{get col names in cluster.}
        \STATE schema $\leftarrow \textsc{getSchema}(\D, \cat, \text{tbls}, \text{cols})$\label{get_clusterschema}\algorithmiccomment{get cluster schemas.}        
        \STATE \ts[c] $\leftarrow \textsc{createDatabase}(\D, \text{tbls}, \text{cols}, \text{schema}, \tau)$ \label{create_db}\algorithmiccomment{add DBs.\ }
     }\ENDFOR

    \RETURN  \ts
\end{algorithmic}
\end{algorithm}
%\resetspace % reset space after algorithmc 

 \begin{figure}[!t]
 \vspace{-0.25cm}
    \resizebox{.99\columnwidth}{!}{\includegraphics[scale=1]{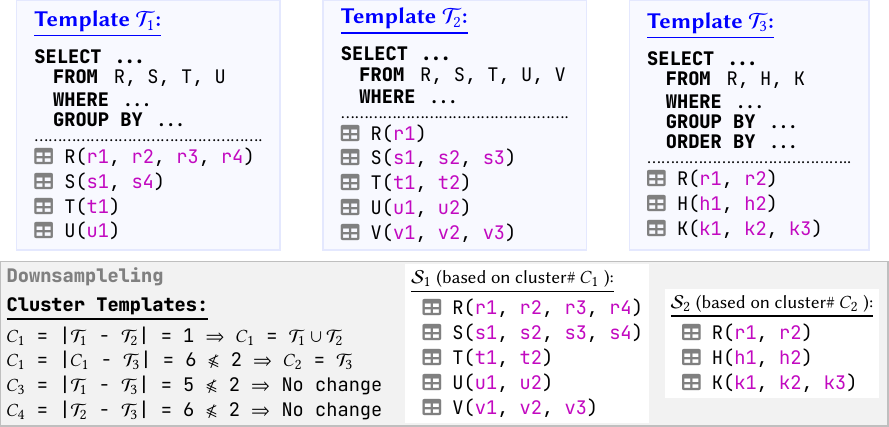}}
    \vspace{-0.35cm}
    \caption{\label{fig:downsampleling-example}Example Template Clustering and Schemas.}
     \vspace{-0.35cm}
\end{figure}

\textbf{Schema Selection:} We first iterate over the list of templates (\qtl) extracted from the workload and auxiliary data catalog information (\cat). At the end, we obtain a subset of the schema relevant for the query workload. Specifically, in Line~\ref{itr:templates}-\ref{get_cols}, we iterate over the templates, and extract all tables and columns accesses in these templates. Next, we prune the original database schema to the subset relevant to these accessed tables, columns, filter indexes, primary keys (PK), and foreign keys (FK) (see Lines~\ref{get_templateschema} and \ref{save_info}).

\textbf{Schema Clustering:} To avoid creating a separate schema per template, we cluster templates by merging overlapping table sets (Line~\ref{templateclustering}). We first cluster templates that access the same tables. Then, we add a template to an existing cluster if it differs by up to two tables. This clustering substantially reduces the number of schemas. 

\begin{example}[Template Clustering and Schema Creation]
Figure~\ref{fig:downsampleling-example} shows an example of template clustering and related schema creation. ReSequel materializes the tables and columns necessary for projection, join, and filter operations. Templates $\qt_1$ and $\qt_2$ share four common tables but differ by one table. Consequently, we merged $\qt_1$ and $\qt_2$ into a single cluster, $C_1$. In the subsequent clustering iteration, we compare the updated cluster $C_1$ with template $\qt_3$. Since the difference between $C_1$ and $\qt_3$ exceeded our (configurable) threshold of two tables, we create a new cluster $C_2$ for $\qt_3$. We also evaluate the pairs ($\qt_1$, $\qt_3$) and ($\qt_2$, $\qt_3$), but because these pairs also exceed the threshold, and the tables were already covered by clusters $C_1$ and $C_2$, no changes are made to the clusters.
For each cluster, ReSequel finally merges the columns of overlapping tables to create a unified set of columns. Additionally, we analyze the original data catalog and incorporate relevant physical design choices (e.g., indexes) into the sample schema.
\end{example}

\textbf{Database Sampling:} 
Downsampling and verification are performed per template and reused across query instances. Hence, their cost is amortized over the workload.
After schema clustering, we generate a diverse set of databases per cluster. For instance, cluster $C_1$ may contain up to $\tau$ databases. This design serves the purpose of reducing risk of validating correctness and performance of queries on a single small sample. In Lines~\ref{itr:cluster}-\ref{create_db}, we extract tables, columns, and the relevant schema for each cluster, then sample the database based on column dependencies. Line~\ref{create_db} (\textsc{createDatabase}) is the most costly step. We first identify the central table in the subset based on FK dependencies, that is the table most connected to others via outgoing or incoming FKs. We then take a uniform random sample of rows from this central table, and follow relationships to sample matching rows from connected tables to prevent the well-known issue of empty results over independent samples \cite{LeisRGK017,GemullaRL08}. Our sampling procedure also accounts for the following challenges:
\begin{itemize}
   \item \emph{Missing Relationships:} Ideally, subset tables preserve FK relationships, but some may be absent due to write-optimization or queries relying on non-key columns. In such cases, we analyze join predicates to infer relationships and leverage feature dependencies collected in the metadata.

   \item \emph{Multiple Central Tables:} If multiple central table candidates exist (e.g., galaxy or snowstorm schemas), and since we generate multiple databases per cluster, we start from these candidates round robin for different samples.

   \item \emph{Sample Selection:} We choose samples that cover diverse data characteristics, e.g., frequent and infrequent values as well as \texttt{NULL}s, to improve robustness across workload patterns.
\end{itemize}

%% file: 6_Rewriting.tex
\section{Query Rewriting}
\label{sec:rewriting}

With the query templates and sampled databases at hand, we are ready to rewrite queries. ReSequel is a prompt-based system that guides an LLM to produce optimized queries using rewrite rules, metadata, and task-specific instructions. Cleaning up syntactic issues (see Figure~\ref{fig:messy_query}) typically requires one or two LLM iterations. However, optimizing queries is more challenging because runtime data is unavailable to the LLM. 
To address this issue, ReSequel constructs prompts in a structured manner: each prompt is derived from the query template and catalog metadata, and targets a single optimization task per iteration. Finally, all candidate rewritten queries are verified through execution, and the best candidate is selected based on its performance on sampled databases.

\subsection{Overall Template Rewriting}
Algorithm~\ref{alg:rewrite} outlines the end-to-end rewriting of a query workload. We assume a streaming model, where each incoming query is processed independently; static workloads are handled by iterating over all queries. To minimize redundant computation, ReSequel employs a two-level cache: a \texttt{Query Cache} for rewritten queries and a \texttt{Template Cache} for validated rewritten templates (Figure~\ref{fig:resequel_arch}).

\textbf{Query Templatization:} For an input query, we first extract its parameterized template and parameter bindings using Algorithm~\ref{alg:templatization}, Line~\ref{get_templateparams}. Subsequently, we attempt query reconstruction by probing the Query Cache to retrieve a previously rewritten query. A cache miss---typical for unseen templates or during warm-up---triggers the LLM-based rewriting pipeline (Lines~\ref{rc_missed}-\ref{invoke_top1}).

\textbf{Database Sampling} As part of metadata management (Line~\ref{build_catalog}), we construct an extended data catalog that augments the native DBMS catalog with column-level statistics and inferred dependencies not explicitly declared in the schema. Additionally, we execute Algorithm~\ref{alg:downsampling} to generate samples for template clusters.

\textbf{Rewriting and Evaluation} For each extracted template, we construct concise, task-specific prompts (Line~\ref{get_prompt}) and submit them in batch to the LLM, producing multiple candidate rewritten templates (Line~\ref{rewrite_template}). 
%
%Since predicate values are masked, these candidates are not directly executable. 
The LLM is used exclusively to generate candidate rewritten templates; correctness and performance decisions are enforced by execution-based validation and selection.
Specifically, we execute the original query and reconstructed queries for each template variant on the sampled database and compare results. Only candidates with identical results are admitted into the \texttt{Template Cache} (Line~\ref{update_tc}). Subsequent invocations benefit from cache hits, and we return the Top-1 candidate as final rewritten query (Line~\ref{invoke_top1}).

%%%%%%%%%%%%%%%%%%%%%%%%%%%%%%%%%%%%%%%%%%%%%%%%%%%%%%%%%%%%%%
%\setspace{6pt} % reduce space after algorithm
\begin{algorithm}[t] \fontsize{7.9}{0}
\caption{\textsc{queryRewrite}$(\Q, \llm,\D,\tau)$}\label{alg:rewrite}
\begin{algorithmic}[1]
  \REQUIRE{Raw Query \Q, LLM \llm, Database \D, Instance Limits $\tau$}
  \ENSURE{New Query \QP}
    % Step A
    \STATE \qt,\p $\leftarrow \textsc{templatization}(q)$ \label{get_templateparams}\algorithmiccomment{get query template and params.}
    \STATE \QP $\leftarrow \textsc{reconstructQuery}(\Q,\qt,\p)$ \label{call_construction}\algorithmiccomment{retrive from cache.}
    \IF[Rewritten Cache missed.]{\QP $= NULL$}  \label{rc_missed}
      \STATE \cat $\leftarrow \textsc{readAndBuildCatalog}(\D)$\label{build_catalog} \algorithmiccomment{build data catalog.}  
      \STATE \ts $\leftarrow \textsc{databaseSampling}(\qt, \D, \cat, \tau)$ 
      \STATE prompt $\leftarrow \textsc{prompt}(\qt, \cat, \llm)$ \label{get_prompt}\algorithmiccomment{get rewrite prompt.}
      \STATE \R $\leftarrow \textsc{submitPromptToLLM}(\text{prompt},\llm) $\label{rewrite_template}\algorithmiccomment{t=\{$t_1,...,t_n$\}.}
      \STATE O $\leftarrow \textsc{runQuery}(\text{\Q},\ts)$ \label{cache_oresults}\algorithmiccomment{execute orig query on sampled DB.} 
      \FOR[iterate over raw rewritten templates.]{$r \in \text{\R}$}{ \label{itr:raw_templates}
          \STATE q $\leftarrow \textsc{buildQuery}(r,\p)$ \label{reconstruct_query}\algorithmiccomment{replace params on template.}
          \IF[all results identical.]{$\textsc{runQuery}(\text{q},\ts)=\text{O}$}  \label{results_eq}
            \STATE $\textsc{TemplateCache} [\qt]\leftarrow \bigcup \text{r}$ \label{update_tc}\algorithmiccomment{update Template Cache.\ } 
          \ENDIF   
        }\ENDFOR 
      \STATE \QP $\leftarrow \textsc{reconstructQuery}(\Q,\qt,\p)$ \label{invoke_top1}\algorithmiccomment{get Top-1 query.\ } 
    \ENDIF   
    \RETURN \QP
\end{algorithmic}
\end{algorithm}
%%%%%%%%%%%%%%%%%%%%%%%%%%%%%%%%%%%%%%%%%%%%%%%%%%%%%%%%%%%%%%

\subsection{Prompt Construction}
\input{tables/prompt.tex}

For guiding the LLM to efficiently generate effective versions of query templates, we aim at global template optimization, where generated versions should cover more than one query. The generated prompts include the masked template, tasks and instructions, example optimization rules, as well as metadata and statistics. 

\textbf{Tasks and Instructions:} Since we do not know upfront which information is needed for the LLM's rewriting process, we base our approach on optimization tasks and instructions. We start from operations of a query template and create tasks for types of optimizations. Table~\ref{tbl:prompt} shows an example of the tasks and instructions, verified against the schema, and used to iteratively prompt the LLM based on the associated examples and metadata.
ReSequel includes instruction prompts based on existing operations in the template and attaches metadata related to these instructions as user messages. Additionally, and in order to allow for exploring a diversity of rewritten queries, we provide examples of possible optimizations as rules. These examples are not limited to the datasets, and the LLM is asked to generate several versions of the template.

%\setspace{6pt} % reduce space after algorithm
\begin{algorithm}[!t] \fontsize{7.9}{0}
\caption{\textsc{reconstructQuery}$(\Q,\qt,\p)$}\label{alg:reconstruct}
\begin{algorithmic}[1]
  \REQUIRE{Query \Q, Query Template \qt, Query Parameters \p}
  \ENSURE{New Query \QP}
    \STATE $\QP \leftarrow \textsc{QueryCache}[\qt, \p]$ \label{read_qc}\algorithmiccomment{read Query Cache.}
    \STATE \R $\leftarrow \textsc{TemplateCache}[\qt]$ \label{read_tc}\algorithmiccomment{read Template Cache.}
    \IF[query cache hit.]{\QP$\neq NULL$}  \label{qc_hit}
           \RETURN $\textsc{buildQuery}(\text{t}, \p)$ \label{get_query_qc}\algorithmiccomment{reconstruct query by cache.} 
    \ELSIF[template cache hit.]{$\R \neq NULL$} \label{tc_hit}     
      \STATE \ts $\leftarrow \textsc{readDatabaseSampling}(\qt)$ \label{get_sample_db}
      \STATE \qtl $\leftarrow [\Q]$ \label{job_list} 
      \FOR[iterate over verified rewritten templates.]{$r \in \text{\R}$}{ \label{itr:versions}
          \STATE q $\leftarrow \textsc{buildQuery}(r,\p)$ \label{build_query}\algorithmiccomment{replace masks with params.}
          \STATE \qtl $\leftarrow  \bigcup \text{q}$ \label{add_q_to_list}
        }\ENDFOR 
      \STATE \QP $\leftarrow \textsc{Top-1}(\text{\qtl})$  \label{get_top1}\algorithmiccomment{rank queries on sample DBs.}
      \STATE $\textsc{QueryCache}[\qt, \p] \leftarrow \QP$ \label{update_qc}\algorithmiccomment{update Query Cache.}
      \RETURN \QP
    \ELSE
      \RETURN $NULL$
    \ENDIF
\end{algorithmic}
\end{algorithm}
%\resetspace % reset space after algorithmc 

\subsection{Query Reconstruction and Caching}
Following LLM-based rewriting, ReSequel invokes Algorithm~\ref{alg:reconstruct} to reconstruct concrete queries by instantiating predicate values and to evaluate their performance.
ReSequel first probes the \texttt{Query Cache}. If a rewritten instance of the same template with similar parameter bindings exists, the system retrieves the cached rewritten query (Line~\ref{read_qc}), instantiates it with the current parameters (Line~\ref{get_query_qc}), and returns the query. Upon a cache miss, ReSequel retrieves the set of verified rewritten templates produced by Algorithm~\ref{alg:rewrite} (Line~\ref{read_tc}).
We then construct a candidate set by instantiating each rewritten template with the original parameter values and adding the original query as a baseline candidate (Lines~\ref{job_list}-\ref{add_q_to_list}). These candidate queries are evaluated on the downsampled database (Line~\ref{get_sample_db}). Execution is performed in parallel with early termination: once a candidate completes, the remaining executions are aborted, preventing excessive overhead from poorly performing queries (Line~\ref{get_top1}). Although this evaluation is conducted on sampled data and may yield different execution plans, our sampling strategy empirically produces near-optimal selections.
Finally, the fastest candidate is inserted into the \texttt{Query Cache} for reuse (Line~\ref{update_qc}) and returned as the rewritten query. If no verified rewritten templates are available, the procedure returns \texttt{NULL}, indicating that no rewrite can be applied.  

\subsection{System Limitations}
ReSequel focuses on query rewriting for read-only workloads and assumes independent query execution. Its database downsampling strategy works well when queries access subsets of tables and columns. It becomes more challenging for queries that span the entire database, where preserving the data distributions is difficult. In addition, queries with side effects, such as user-defined functions that perform updates, are not explicitly handled. 
%These cases may lead to undetected correctness or performance issues.

%% file: tables/prompt.tex
\begin{table*}[ht]
\vspace{-0.4cm}
    \centering
    \caption{The prompts used to generate the optimized template version are based on tasks and operations. In addition to the prompt, we also encode the schema (e.g., relevant tables, columns, and indexes) as well as statistics in the requests. }
    \vspace{-0.4cm}
    \label{tbl:prompt}
    \resizebox{0.99\textwidth}{!}{
        \addtolength{\tabcolsep}{-.1em}	
        \begin{tabular}{c|p{13cm}|p{10cm}}
        \toprule
        \textbf{V\#}& \textbf{Prompt Content (Template SQL Script + Projected Schema + Metadata + Examples)} & \textbf{LLM Recommend Example (Implemented Functions + New Template)} \\ \midrule 
        $V1$              & \texttt{Rewrite SQL query for optimization, possibly using \textbf{data skipping through prefiltering}}.& \texttt{\textblue{CREATE INDEX} \textbperpal{trgm\_idx} \textblue{ON} \textbperpal{name} \textblue{USING} gin (\textbperpal{name} gin\_trgm\_ops);} \\ \chline
        $V2$              & \texttt{Rewrite SQL query for optimization by \textbf{recommending appropriate data structures}.}& \texttt{\textblue{SET LOCAL} enable\_nestloop = \textbperpal{off}; \textblue{SET LOCAL} enable\_hashjoin = \textbperpal{on};}  \\ \chline
        $V3$              & \texttt{\textbf{Implement functions} to optimize queries involving \textbf{equality (=) conditions}.} & \texttt{\textblue{CREATE INDEX} \textbperpal{idx\_hash\_role} \textblue{ON} \textbperpal{role\_type} \textblue{USING HASH} (role);} \\ \chline
        $V4$              & \texttt{Rewrite the SQL query based on \textbf{join types and filter operations} ($p1$).} & \texttt{Push filters to scans, replace cross joins with explicit INNER JOINs and use EXISTS for many‑to‑many links.} \\ \chline
        $V5$              & \texttt{Optimize the query by \textbf{implementing a join cardinality estimation formula} ($p2$).} & \texttt{Formula: (\textbperpal{rows\_A} * \textbperpal{rows\_B}) / max(\textbperpal{distinct\_A}, \textbperpal{distinct\_B}).} \\ \chline
        $V6$              & \texttt{Enables \textbf{probabilistic data structures} for highly efficient Joins($p3$).} & \texttt{\textblue{CREATE EXTENSION IF NOT EXISTS} bloom;} \\ \chline
        $V7$              & \texttt{Rewrite the template to \textbf{optimize subsequent filtering}.} & \texttt{Optimized with a CTE for the filtered table and EXISTS clauses.}\\  \chline
        $V8$              & \texttt{\textbf{Candidate Key Identification}: Analyzes the provided schema statistics to identify columns where the count of distinct values.} & \texttt{\textblue{CREATE FUNCTION} find\_candidate\_keys(\textbperpal{table\_name} \textblue{TEXT})} \\ 
        % $V9$              & The optimized query uses a JOIN, processing the match once, with DISTINCT to avoid duplicates.& \\
        \bottomrule
    \end{tabular}
}  
\vspace{-0.1cm}
\end{table*}

% Orig Query: 
% SELECT customer_id
% FROM Customers c
% WHERE EXISTS (
%     SELECT 1
%     FROM Orders o
%     WHERE o.customer_id = c.customer_id
%     AND o.amount > 1000
% );

% new query:
% SELECT DISTINCT c.customer_id
% FROM Customers c
% JOIN Orders o ON c.customer_id = o.customer_id
% WHERE o.amount > 1000;

% The optimized query uses a JOIN, processing the match once, with DISTINCT to avoid duplicates. 

%% file: 7_Experiments.tex
\begin{figure*}[!t]
      \centering   
    \includegraphics[scale=0.5]{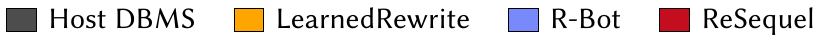}\vspace{-0.2cm}

    \subfigure[\label{fig:overall_1pg}PostgreSQL]{  
        \includegraphics[scale=0.66]{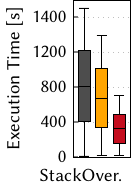}
        \includegraphics[scale=0.66]{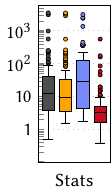}
        \includegraphics[scale=0.66]{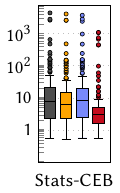}
        \includegraphics[scale=0.66]{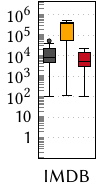}
        \includegraphics[scale=0.66]{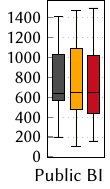}
    }
    \includegraphics[scale=0.42]{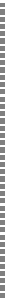}
    \subfigure[\label{fig:overall_1duckdb}DuckDB]{  
        \includegraphics[scale=0.66]{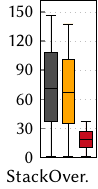}
        \includegraphics[scale=0.66]{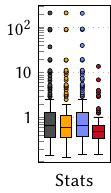}
        \includegraphics[scale=0.66]{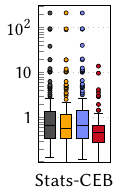}
        \includegraphics[scale=0.66]{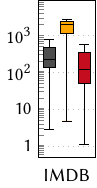}
        \includegraphics[scale=0.66]{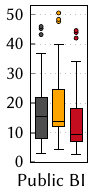}
    }
 \includegraphics[scale=0.42]{figures/img/fig_delimiter}
    \subfigure[\label{fig:overall_1mysql}MySQL]{  
        \includegraphics[scale=0.66]{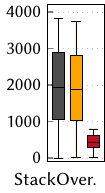}
        \includegraphics[scale=0.66]{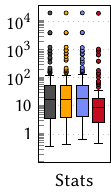}
        \includegraphics[scale=0.66]{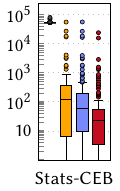}
    }
    
    \vspace{-0.6cm}  
    \caption{Overall Performance Impact on Various DBMSs Across 5 Datasets (LLM: Gemini-2.5-pro; 10\% Downsampling).}
    \label{fig:overall_1}
    \vspace{-0.3cm}    
\end{figure*}

\section{Experiments}
\label{sec:experiments}

We evaluate ReSequel on a diverse set of benchmarks, including real-world datasets, across multiple DBMSs and LLMs. 

\subsection{Experimental Setup}
\textbf{Implementation Details:} ReSequel's rewriting, verification, and top-1 selection is implemented in Python, whereas the downsampling is written in C++ (which reduced the overhead by up to 15x). For rewriting, we use Google AI Studio (Gemini-2.5-pro)~\cite{gemini} and Groq Cloud (OSS-120B)~\cite{groq}, whose  latency can vary substantially (across LLMs and peak times).
To build the data catalog, we extract the host DBMS's catalog, and execute SQL data profiling queries to obtain statistics. ReSequel currently supports three DBMS dialects: PostgreSQL, MySQL, and DuckDB.

\textbf{HW/SW Environment:} We ran all experiments on a VM with an Intel CPU (32 vCPUs) and 148\,GB DDR4 RAM. The software stack comprises Ubuntu 22.04, Python 3.10, C++ 17, PostgreSQL 17.1, MySQL 8.0.43, and DuckDB 1.4.0. We tuned PostgreSQL using PGTune~\cite{PGTune} for general OLAP workloads. For MySQL, we configured InnoDB with \texttt{buffer\_pool\_size=100GB}. DuckDB requires no special tuning. All DBMSs were run with multi-threading enabled.

\textbf{Benchmarks:} We evaluate ReSequel on the TPC-H (sf=10)~\cite{tpch}, DSB (sf=100)~\cite{DSB}, Stats/Stats-CEB~\cite{CEDBMS}, Public BI Benchmark~\cite{publicBI} (top 100 longest-running queries), Join Order Benchmark (JOB)~\cite{LeisGMBK015}, IMDB Full~\cite{flowloss}, and StackOverflow~\cite{SQlStorm}. Our objectives are to assess the performance benefits of LLM-assisted rewriting, and compare template- and query-based rewriting. We use three workload groups: \emph{Group 1}, where templates and queries are identical except for masked parameters (TPC-H, Stats/Stats-CEB, DSB, Public BI); \emph{Group 2}, where $\approx15\%$ of queries map to the same template (JOB); and \emph{Group 3}, subsets of \numprint{1192} queries from SQLStorm for StackOverflow, and \numprint{13646} from IMDB, with high overlap and few templates. Table~\ref{tbl:datasets} summarizes the data and workload characteristics.

\input{tables/datasets.tex}

\textbf{Baselines:} Furthermore, we consider three types of baselines: (1) a \emph{Host DBMS systems}, where we run unmodified queries; (2) two \emph{LLM-based systems (LLM-R2~\cite{LLMR2} and R-Bot~\cite{RBot})} that use LLMs to apply rules; and (3) a \emph{Non-LLM-based, learned-rewrite system (LearnedRewrite (\textbf{LR})~\cite{LearnedRewrite})}, which rewrites queries using a cost estimation model and Monte Carlo Tree Search to identify the best query. 
To verify the equivalence of ReSequel-rewritten queries, we used SQLSolver~\cite{DingWYZXCPL23} as a baseline for ReSequel downsampling verification.  
R-Bot and LLM-R2 are methods that rewrite individual queries and therefore cannot handle workloads with many queries. In addition, LLM-R2 requires training on query plans, which is not available for all benchmarks. Therefore, we ran R-Bot for Groups 1 and 2 of workloads, and LLM-R2 for JOB, TPC-H, and DSB.

\subsection{End-to-End Experiments}
We evaluate both the overall and individual query performance.

\begin{figure*}[!t]
    \vspace{-0.35cm}
    \centering   
    \includegraphics[scale=0.5]{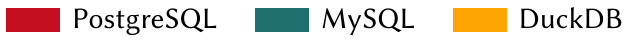}\vspace{-0.25cm}
    \resizebox{.99\textwidth}{!}{
        \subfigure[\label{fig:individual_1_a}StackOverflow]{\includegraphics[scale=1]{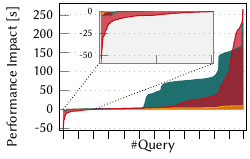}}
        \hfill   
        \subfigure[\label{fig:individual_1_b}Stats]{\includegraphics[scale=1]{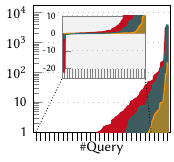}}
        \hfill   
        \subfigure[\label{fig:individual_1_c}Stats-CEB]{\includegraphics[scale=1]{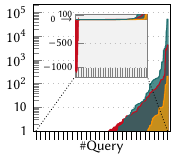}}
        \hfill   
        \subfigure[\label{fig:individual_1_d}Public BI]{\includegraphics[scale=1]{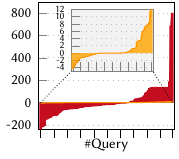}}
        \hfill   
        \subfigure[\label{fig:individual_1_e}IMDB]{\includegraphics[scale=1]{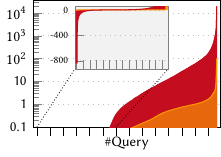}}
    }
    \vspace{-0.6cm}  
    \caption{Individual Performance Impact on Workloads of Five Datasets (Overall = Figure~\ref{fig:overall_1}, LLM = Gemini-2.5-pro).}
    \label{fig:individual_1}
    \vspace{-0.4cm}    
\end{figure*}

\begin{figure*}[!t]
    \centering  
    \resizebox{.99\textwidth}{!}{ 
        \subfigure[JOB]{\includegraphics[scale=.91]{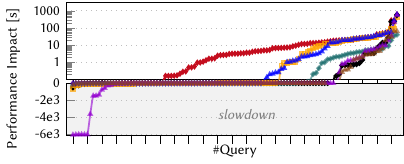}}
        \hfill    
        \subfigure[TPC-H]{\includegraphics[scale=.81]{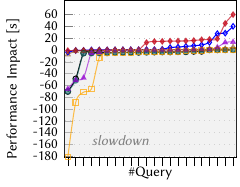}}
        \hfill
        \subfigure[DSB]{\includegraphics[scale=.91]{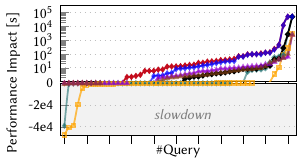}}\hspace{0.5cm}        
        \includegraphics[scale=.8]{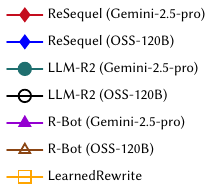}
    }

    \vspace{-0.6cm}  
    \caption{Individual Performance Impact on Workloads of Three Datasets (Overall = Figure~\ref{fig:overall_2}, DBMS = PostgreSQL).}
    \label{fig:individual_2}
    \vspace{-0.4cm}    
\end{figure*}

\begin{figure*}[!t]
    \centering    
    \resizebox{.99\textwidth}{!}{ 
        \includegraphics[scale=0.9]{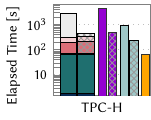} 
        \includegraphics[scale=0.9]{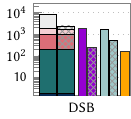}
        \includegraphics[scale=0.9]{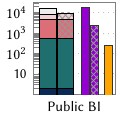}
        \includegraphics[scale=0.9]{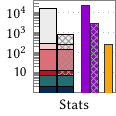}
        \includegraphics[scale=0.9]{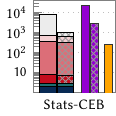}            
        \includegraphics[scale=0.9]{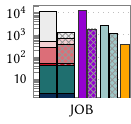}
        \includegraphics[scale=0.9]{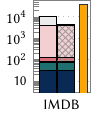}
        \includegraphics[scale=0.9]{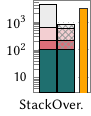} \hspace{0.1cm}          
        \includegraphics[scale=0.33]{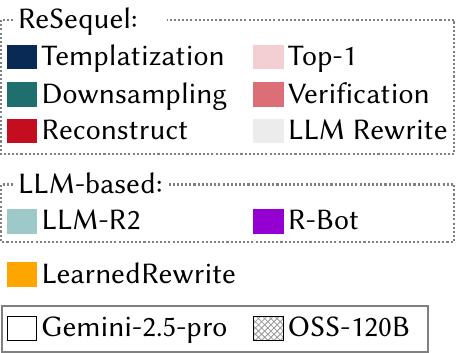}     
    }
    \vspace{-0.4cm}  
    \caption{\label{fig:rewrite_exe_time} Time Breakdown of Rewriting Across Eight Datasets (Downsampling Rate = 10\%, DBMS = PostgreSQL).}
    \vspace{-0.3cm}    
\end{figure*}

\begin{figure}[!t]
    \centering   
    \includegraphics[scale=0.47]{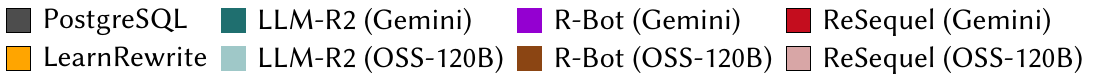}\vspace{-0.15cm}

    \subfigure[\label{fig:overall_2_job}JOB]{\includegraphics[scale=0.73]{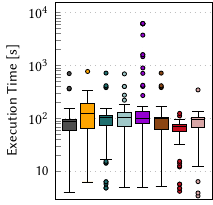}}
    \hfill
    \subfigure[\label{fig:overall_2_tpch}TPC-H]{\includegraphics[scale=0.73]{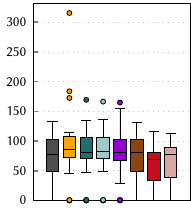}}
    \hfill
    \subfigure[\label{fig:overall_2_dsb}DSB]{\includegraphics[scale=0.73]{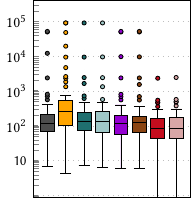}}

    \vspace{-0.5cm}  
    \caption{Performance Impact on Three Datasets.}
    \label{fig:overall_2}
    \vspace{-0.45cm}    
\end{figure}

\input{tables/rewrite_ratio.tex}
\input{tables/sqlsolver.tex}

\textbf{Rewriting Robustness:} To illustrate the effectiveness of the baselines in modifying workload queries, Table~\ref{tbl:modified_ratio} reports the ratio of successfully modified queries. This ratio is measured on PostgreSQL; the results for MySQL and DuckDB differ by at most $\approx1\%$. The LLM-based baselines and LearnedRewrite show limited success in rewriting workloads of queries while preserving query equivalence, and they also fail to support workloads with many queries. Here, we selected five datasets for which the baselines exhibited relatively high overlap. 
Additionally, Table~\ref{tbl:sqlsolver} summarizes SQLSolver performance (outputs: Equal, Unknown). The results show most LLM-generated queries are unsupported by equivalence checkers like SQLSolver because they often fall outside the recognized AST patterns. Additionally, execution time can exceed four days for some workloads (e.g., Public BI), rendering it substantially slower than ReSequel's downsampling-based verification.

\textbf{Workload Performance Impact:} Figure~\ref{fig:overall_1} (datasets from Groups 1 \& 3) presents the overall performance impact of ReSequel's rewriting. First, on StackOverflow and IMDB, ReSequel merges many queries into a small number of templates (StackOverflow: 60 templates for 1,190 queries; IMDB: 79 templates for 13,646 queries). The overall workload execution time improved by up to 2.45x on PostgreSQL, 4.81x on MySQL, and 3.95x on DuckDB. The StackOverflow workload---generated using GPT-4o-mini---contains many messy queries, where existing rewrite systems often struggle. Similarly, the IMDB workload includes redundant join orders, which makes cost-based optimization difficult. LearnedRewrite rewrites queries individually and achieves limited gains, as its Calcite rules also struggle with such messy queries. In contrast, ReSequel achieves substantial improvements through rewrites such as early string filtering, data-structure hints, dynamic join ordering, redundant join elimination, and subquery simplification and reuse.
Second, for Stats, Stats-CEB, and Public BI, only literal values were masked. On Stats and Stats-CEB, ReSequel improved execution time by 5.89x/4x on PostgreSQL, 14.85x/16x on DuckDB, and 3.3x on MySQL (Stats-CEB). These gains stem from generating up to 20 alternative query variants per input query, breaking the original query into parts (some operating on data and others on the catalog), and removing redundancies in aggregation queries (e.g., redundant \texttt{GROUP BY} clauses). On Public BI, ReSequel shows limited improvement. The dataset is large (386 GB), which restricts downsampling diversity and weakens Top-1 selection. Many queries also use range predicates (where masking leads to over-generalized filters) and self-joins (where rewrite hints occasionally degrade performance).
R-Bot successfully rewrites 12\% of the Stats/Stats-CEB workload, but with negative performance impact. Although R-Bot is able to run on the Public BI workload, it incurs substantial overhead and fails to successfully rewrite even a single query (see Table~\ref{tbl:modified_ratio}).

\textbf{Query Performance Impact:} Figures~\ref{fig:individual_1} and \ref{fig:individual_2} compare the performance of rewritten and original queries. Templatization consistently improves performance, with most queries experiencing speedups. These results show only the performance of rewritten queries, whereas the rewriting overheads are discussed in Section~\ref{sec:micro_bench}. Although a few queries experience slowdowns, their impact on overall performance is negligible. Several high-cost queries of StackOverflow and IMDB achieve substantial improvements. Public BI remains the only workload with limited gains. However, many queries on DuckDB still improve due to reduced numbers of intermediates. In contrast, LLM-based systems and LearnedRewrite improve only few queries and often incur slowdowns.

\textbf{Impact of Different LLMs:} Figure~\ref{fig:overall_2} shows the execution time of ReSequel on JOB, TPC-H, and DSB using two LLMs (Gemini and OSS-120B). 
On JOB, PostgreSQL already performs well for most queries, but several complex queries (e.g., 20a-20c) contain redundant joins (14 joins over 10 tables) and benefit from ReSequel’s rewrites (Figure~\ref{fig:overall_2_job}). These improvements come from single-table subqueries combined with \texttt{INTERSECT}. LearnedRewrite yields small gains through join reordering, while LLM-R2 often produces suboptimal plans with very large intermediates (>200 GB). Most JOB queries rewritten with Gemini exhibit slowdowns. ReSequel achieves speedups of 6.18x, 5.64x, 5.76x, and 50x over LearnedRewrite, PostgreSQL, LLM-R2, and R-Bot.
For TPC-H (Figure~\ref{fig:overall_2_tpch}), the workload diversity is low and thus, rewriting opportunities are limited. ReSequel still improves performance by 1.14x over PostgreSQL. LearnedRewrite and LLM-based systems show minimal or negative gains. Differences from previously reported results are due to the underlying LLMs (Gemini and OSS-120B instead of GPT-4), indicating strong sensitivity.
The DSB workload (Figure~\ref{fig:overall_2_dsb}) is skewed and highly correlated. Here, LLM-R2 and R-Bot perform worse than with GPT-4. ReSequel remains robust by generating multiple candidate rewrites per query. This strategy increases the chances of at least one variant showing improvements. Overall, ReSequel achieves speedups of 38.6x, 22x, 38.56x, and 22x.

\vspace{-0.3cm}
\subsection{Micro Benchmarks}
\label{sec:micro_bench}
In order to further understand these end-to-end results and the source of improvements, we conduct a variety of micro-benchmarks. 

\textbf{Time Breakdown of Rewriting:} Figure~\ref{fig:rewrite_exe_time} shows the total and per-step rewriting time of ReSequel and baselines (in log scale) across all datasets on PostgreSQL using Gemini-2.5-pro and OSS-120B. 
\ding{202}, workload templatization is lightweight and negligible. 
\ding{203}, building the extended data catalog is the main bottleneck; for fairness, timing reflects the full dataset even though smaller versions would suffice. 
\ding{204}, the LLM rewrites templates; generation time scales with template count, and Gemini is slower than OSS-120B due to its thinking mode. 
\ding{205}, workload reconstruction replaces placeholders with original values; runtime depends on query complexity, and topological reconstruction handles duplicate keys. 
\ding{206}, during verification, the original query and up to $N$ reconstructed variants per template are executed on sampled databases; total time grows with template count and $N$. 
\ding{207}, $Top\text{-}1$ selection chooses the fastest candidate per query with parallel execution, early stopping, and caching.
Differences on other DBMSs stem mostly from catalog construction, verification, and top-1 selection.

\begin{figure*}[!t]
    \vspace{-0.35cm}
    \centering  
    \includegraphics[scale=0.5]{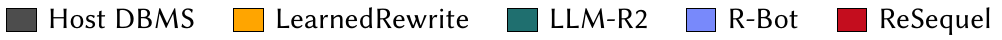}

    \resizebox{.99\textwidth}{!}{ 
        \includegraphics[scale=0.9]{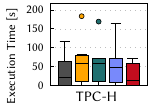} 
        \includegraphics[scale=0.9]{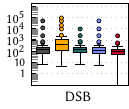}
        \includegraphics[scale=0.9]{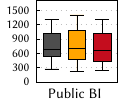}
        \includegraphics[scale=0.9]{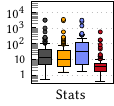}
        \includegraphics[scale=0.9]{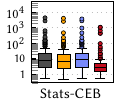}            
        \includegraphics[scale=0.9]{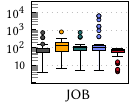}
        \includegraphics[scale=0.9]{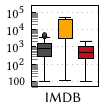}
        \includegraphics[scale=0.9]{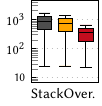}         
  
    }
    \vspace{-0.4cm}  
    \caption{\label{fig:specific_queries} Impact of Queries with Messy Characteristics Across Datasets (Gemini-2.5-pro \& PostgreSQL).}
    \vspace{-0.25cm}    
\end{figure*}

\begin{figure*}[!t]
    \centering     
    \resizebox{.7\textwidth}{!}{\includegraphics[scale=1]{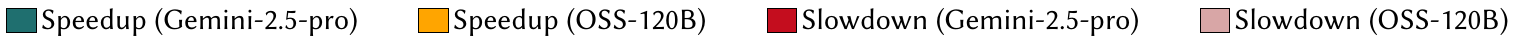}} \vspace{-0.2cm}    

    \resizebox{.99\textwidth}{!}{ 
    \subfigure[Ratio=10\%]{\includegraphics[scale=0.59]{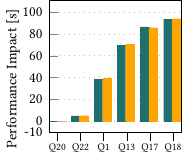}}
    \subfigure[Ratio=20\%]{\includegraphics[scale=0.59]{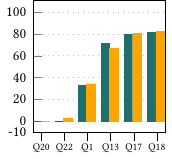}}
    \subfigure[Ratio=30\%]{\includegraphics[scale=0.59]{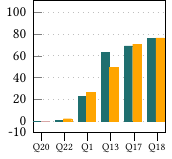}}
    \subfigure[Ratio=40\%]{\includegraphics[scale=0.59]{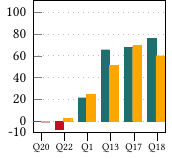}}
    \subfigure[Ratio=50\%]{\includegraphics[scale=0.59]{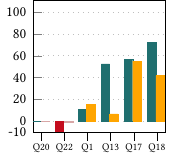}}
    \subfigure[Ratio=60\%]{\includegraphics[scale=0.59]{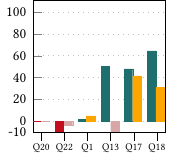}}
    \subfigure[Ratio=70\%]{\includegraphics[scale=0.59]{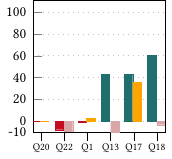}}
    \subfigure[Ratio=80\%]{\includegraphics[scale=0.59]{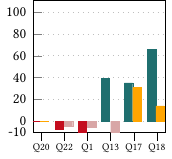}}
    \subfigure[Ratio=90\%]{\includegraphics[scale=0.59]{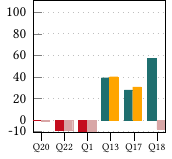}}
    \subfigure[Ratio=100\%]{\includegraphics[scale=0.59]{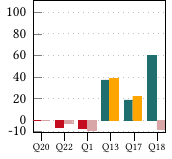}}
    }
    \vspace{-0.5cm}  
    \caption{Individual Query Performance Impact Debugging with Variety of Downsampling Ratios.}
    \label{fig:tpch_s10_impact}
    \vspace{-0.25cm}    
\end{figure*}

\begin{figure}[!t]
%\vspace{-0.15cm}
    \resizebox{.99\columnwidth}{!}{\includegraphics[scale=1]{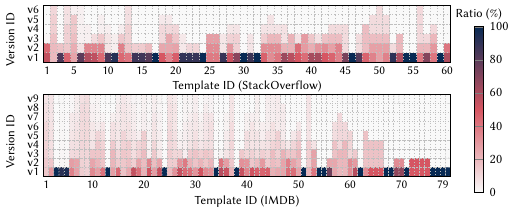}}
    \vspace{-0.5cm}
    \caption{\label{fig:cache_hit}Number of Template and Query Cache Hits.}
     \vspace{-0.25cm}
\end{figure}

\begin{figure}[!t]
    \centering \vspace{-0.15cm}    
    \resizebox{.45\columnwidth}{!}{\includegraphics[scale=1]{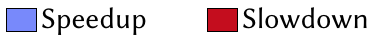}} \vspace{-0.25cm}    

    \subfigure[\label{fig:so_template_impact_a}|Queries|=86]{\includegraphics[scale=0.9]{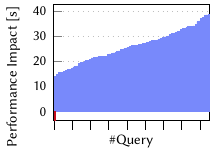}}
    \subfigure[\label{fig:so_template_impact_b}|Queries|=46]{\includegraphics[scale=0.9]{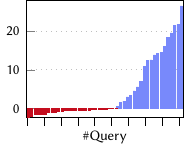}}
    \subfigure[\label{fig:so_template_impact_c}|Queries|=12]{\includegraphics[scale=0.9]{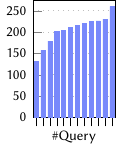}}

    \vspace{-0.5cm}  
    \caption{Individual Query Performance Impact across Three StackOverflow Templates (DBMS: PostgreSQL; LLM: Gemini-2.5-pro; Downsampling Ratio = 10\%).}
    \label{fig:so_template_impact3}
    \vspace{-0.35cm}    
\end{figure}

\textbf{Number of Cache Misses:} ReSequel uses template and query caches to reduce the number of LLM calls and amortize the top-1 selection. For datasets in Groups 1 and 2, all queries and templates result in 100\% cache misses. For StackOverflow and IMDB, which have a large number of queries, only a few templates are rewritten. Figure~\ref{fig:cache_hit} shows the frequency of selecting rewritten template versions during workload execution. For example, in IMDB, for template IDs 3, 4, and 5, all queries select version \#1.  

\textbf{Template-based Performance:} We next evaluate the impact of template-based rewriting using Top-1 selection using three templates from StackOverflow. Figure~\ref{fig:so_template_impact_a} shows the template with the most merged queries, Figure~\ref{fig:so_template_impact_b} represents a mid-sized template, and Figure~\ref{fig:so_template_impact_c} the one with fewest merged queries. ReSequel performs well at both ends of this spectrum. Only one query shows a minor slowdown, while most merged queries improve. In contrast, for the mid-sized template, nearly half of the queries slow down due to aggregations. Here, Top-1 selection on a 10\% sampled database fails to identify the optimal variant. Overall, these results indicate room for improvement. However, we observe no direct correlation between the degree of templatization (i.e., number of merged queries) and the performance of rewritten queries.

\begin{figure}[!t]
    \resizebox{.99\columnwidth}{!}{\includegraphics[scale=1]{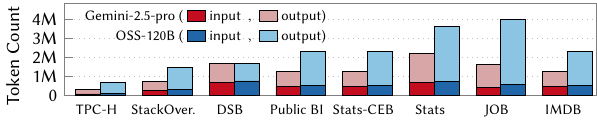}}
    \vspace{-0.4cm}
    \caption{\label{fig:cost}{Costs in terms of Number of LLM Tokens for Rewriting all Query Templates of the Different Benchmarks.}}
     \vspace{-0.35cm}
\end{figure}

\begin{figure}[!t]
    \resizebox{.99\columnwidth}{!}{\includegraphics[scale=1]{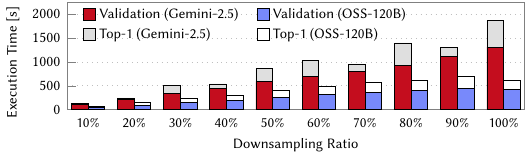}}
    \vspace{-0.45cm}
    \caption{\label{fig:toch_verify_top1}Validation and Top-1 Selection Overhead with Different Downsampling Ratios (TPC-H SF10; PostgreSQL).}
     \vspace{-0.35cm}
\end{figure}

\textbf{Rewriting Costs:} LLM costs scale with the number of input and output tokens. Figure~\ref{fig:cost} reports the tokens for generating ten variants per template. To exceed an LLM’s single-call output limit, we chain multiple requests, which increases token usage due to repeated instructions. For both LLMs, ReSequel uses identical prompts. Output tokens vary by model and include generated SQL queries, explanations, and hidden reasoning tokens. Across eight datasets, ReSequel consumes \numprint{6142842} tokens with Gemini and \numprint{12085412} tokens with OSS-120B. In contrast, baselines that invoke the LLM per query are less efficient. For example, rewriting \numprint{1190} StackOverflow queries with Gemini requires over 14 million tokens.

\textbf{Specific Messy Query Characteristics:} In Figure~\ref{fig:messy_query}, we highlight the fractions of messy query characteristics per dataset using red bars (e.g., Stats exhibits 92\% ambiguous joins). Figure~\ref{fig:specific_queries} shows the corresponding labeled characteristics. We select a subset of queries from each workload and execute them. Compared with the host DBMS and baselines, ReSequel effectively overcomes issues caused by messy queries and outperforms nearly all baselines.

\input{tables/tpch_downsampling_verify_error.tex}

\textbf{Downsampling Ratios:} To study the effect of sampling on validation and Top-1 selection, we use TPC-H SF10 and vary the downsampling ratio from 10\% to 100\%. We analyze six queries: \{Q13, Q17, Q18\} with long execution times and \{Q1, Q20, Q22\} with performance degradation at 10\% sampling ratio. Figure~\ref{fig:tpch_s10_impact} shows the performance impact, and Figure~\ref{fig:toch_verify_top1} shows the validation results. We evaluate two LLMs and observe that Gemini is substantially slower. However, as shown in Table~\ref{tbl:verify_failed_count}, Gemini produces more successfully validated variants than OSS-120B. In contrast, most variants generated by OSS-120B show minimal or no performance impact, leading to faster validation and Top-1 selection. We always include the original query among the candidates, which reduces the risk of performance regressions. OSS-120B achieves weaker runtime improvements due to fewer validated variants.

\begin{figure}[!t]
    \vspace{-0.35cm}
    \resizebox{.99\columnwidth}{!}{\includegraphics[scale=1]{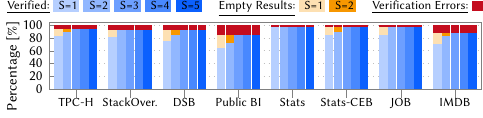}}
    \vspace{-0.45cm}
    \caption{\label{fig:downsampleling_verify}Verification Rates Across Different Sample Sizes (\small{Verification Errors: incorrect LLM rewrites + reconstruction errors}).}
     \vspace{-0.5cm}
\end{figure}

\textbf{Impact of \#Samples:} To study the impact of the number of samples on query verification, for each workload, we generate 1-5 samples, execute the workload, and examine the queries. Figure~\ref{fig:downsampleling_verify} shows the ratio of verified queries, empty results (which are difficult to verify), and errors, i.e., unverified queries (w/ Gemini and PostgreSQL). The queries with errors were identified during the verification stage and simply deemed invalid variants. These results show that, in most cases, two (sometimes three) samples are sufficient to correctly verify the entire workloads.

\textbf{Impact of Rewrite Categories:} To study the sources of rewrite guidance in ReSequel, we performed an ablation study over the three rule categories: (1) \emph{Calcite rules}, i.e., generic rewrite rules from Apache Calcite; (2) \emph{metadata-derived rules}, i.e., rewrite guidance from schema information and workload metadata; and
(3) \emph{instruction-by-example rules}, i.e., manually curated rewrite examples.
We keep the generation, verification, caching, and top-1 selection fixed, while varying only the rule guidance to the LLM. Figure~\ref{fig:ab-a} shows the distribution of rule categories applied by the LLM. Additionally, we decomposed every JOB query into three queries for the different rule categories.
Figures~\ref{fig:ab-b} and \ref{fig:ab-c} show that overall, ReSequel achieves a 5.64x speedup over PostgreSQL. With Calcite rules only, the speedup decreases to 4.4x, indicating that standard algebraic rewrites remain important.
Metadata-derived rules only achieve a 2x overall speedup. These rules have the largest impact when transformations depend on schema semantics, relational properties, or workload context. However, in some cases, they negatively affect performance.
Example-based rules primarily improve performance for long-running queries and queries requiring multiple transformations. In some cases, they introduce greater slowdowns.
Overall, each category exhibits substantial performance variability, but applied together, these---sometimes overlapping---rules are a powerful  augmentation for LLM guidance.

%% file: tables/datasets.tex
\begin{table}[!t]
%\vspace{-0.35cm}
    \centering
    \caption{Used Datasets and their Workload Characteristics.}
    \vspace{-0.4cm} 
    \label{tbl:datasets}
    \resizebox{0.99\columnwidth}{!}{
		%\small 
    \setlength\tabcolsep{10pt}
        %\addtolength{\tabcolsep}{.5em}	
        \begin{tabular}{cccc|c}
        \toprule
        \textbf{Dataset}& \textbf{\# Tables} & \textbf{Workload} & \textbf{\# Queries} &\textbf{\# Templates}\\ \midrule 
        TPC-H           & 8                  & TPC-H (sf=10)     & 22                  & 22 \\
        DSB             & 25                 & DSB (sf=100)      & 52                  & 52 \\        
        Public BI       & 83                 &                   & 100                 & 100 \\
        Stats/Stats-CEB & 8                  &                   & 146                 & 146 \\
        \midrule
				JOB             & 21                 &                   & 113                 & 95 \\                  
        \midrule
        StackOverflow   & 8                  &  SQLStorm         & 1,192               & 60 \\
        IMDB            & 21                 &                   & 13,646              & 79 \\ 				
        \bottomrule
    \end{tabular}
}
\vspace{-0.35cm}  
\end{table}

%% file: tables/rewrite_ratio.tex
\begin{table}[!t]
    \centering
    \caption{Ratio of Rewritten Queries (Gemini \& PostgreSQL).}
    \vspace{-0.4cm} 
    \label{tbl:modified_ratio}
    \resizebox{0.99\columnwidth}{!}{
		%\small \
        \setlength\tabcolsep{3pt}
        %\addtolength{\tabcolsep}{.5em}	
        \begin{tabular}{c|cccccccc}
        \toprule
        \textbf{Baeline}  & \textbf{TPC-H} & \textbf{DSB} & \textbf{Public BI} & \textbf{Stats} & \textbf{Stats-CEB} & \textbf{JOB} & \textbf{StackOverflow} & \textbf{IMDB} \\ \midrule
        LLM-R2            &95\%  &61\%  &N/A  &N/A   &N/A   &92\%    &N/A    &N/A \\ \chline
        R-Bot             &95\%  &5\%   &0\%  &15\%  &15\%  &60\%    &N/A    &N/A \\ \chline
        LR                &86\%  &63\%  &72\% &12\%  &12\%  &74\%    &75\%   &37\% \\ \midrule
        \textbf{ReSequel} &100\% &100\% &100\%&100\% &100\% &100\%   &100\%  &100\%	\\ 
        \bottomrule
    \end{tabular}
}
\vspace{-0.35cm}  
\end{table}

%% file: tables/sqlsolver.tex
\begin{table}[!t]
    \centering
    \caption{Ratio \& Verification Time of SQLSolver (PostgreSQL).}
    \vspace{-0.4cm} 
    \label{tbl:sqlsolver}
    \resizebox{0.99\columnwidth}{!}{
        \setlength\tabcolsep{3pt}  
        \begin{tabular}{l|cccccccc}
        \toprule
                       & \textbf{TPC-H} & \textbf{DSB} & \textbf{Public BI} & \textbf{Stats} & \textbf{Stats-CEB} & \textbf{JOB} & \textbf{StackOverflow} & \textbf{IMDB} \\ \midrule
        Equal [\%]     &27  &0   & 0  & 8   &7   &0    &0    &0 \\ \chline
        %Not Equal [\%] &9   &0   &0   &5  &6   &0    &0    &0 \\ \chline
        Unknown [\%]   &73  &100 &100 &92   &93    &100  &100  &100 \\ \midrule
        \textbf{Exe. Time [min]} &72.5 &\numprint{1142} &\numprint{5764} &2.1 &1.7  &0.6  &3.7 & 59.6	\\ 
        \bottomrule
    \end{tabular}}
\vspace{-0.25cm}  
\end{table}

%% file: tables/tpch_downsampling_verify_error.tex
\begin{table}[!t]
    \centering
    \caption{Count of Verified/Failed Rewritten Queries.}
    \vspace{-0.4cm}
    \label{tbl:verify_failed_count}
    \resizebox{\columnwidth}{!}{
        \addtolength{\tabcolsep}{-.0em}	
        \begin{tabular}{c|cccccc|c}
        \toprule
        \textbf{LLM}& \textbf{Q1} & \textbf{Q13} & \textbf{Q17} & \textbf{Q18} & \textbf{Q20} & \textbf{Q22} & \textbf{Total (Verified/Failed)} \\ \midrule 
         Gemini-2.5 & 8/12 & 18/2 & 17/3 & 20/0 & 10/10 & 14/6 & 120 (87+33)\\
         OSS-120B   & 9/11 & 16/4 & 15/5 & 18/2 & 10/10 & 2/18 & 120 (70+50)\\
        \bottomrule
    \end{tabular}
}  
\vspace{-0.35cm}
\end{table}

%% file: 8_RelatedWork.tex
\vspace{-0.3cm}
\section{Related Work}

% This section reviews prior work on query rewriting, optimization, LLM-based transformation, and query verification.

\textbf{Query Rewrite Systems:} Early work on query rewriting dates back to the 1990s. An example is the Starburst rewrite system~\cite{PiraheshHH92,PiraheskLH97}, which applies heuristic rules through a dedicated rule engine. Such systems support rewrites including common subexpression elimination, DISTINCT push-down and pull-up, predicate push-down, and subquery unnesting into joins. Subsequent work extended rule-based rewriting to new data models and queries, such as object-oriented queries, XML~\cite{KrishnaprasadLMWAK04,GodfreyGHMZ09}, and text~\cite{BaoKL12}.
More recently, similar rule-based rewriting is used in ML systems~\cite{BoehmBERRSTT14} and DNN frameworks~\cite{JiaTWGZA19,FangSW020}. In this setting, approaches such as super-optimization~\cite{JiaPTWZA19} and sum-product optimization~\cite{ElgamalLBETRS17,WangHSHL20} automatically derive new rewrites from algebraic operator properties.
In contrast to these general-purpose rewrite systems, we focus on LLM-based, whole-query rewriting of a workload of queries.
Like other LLM-based systems, GenRewrite~\cite{liu2024query} is a multi-stage rewriting framework. It selects a subset of workload queries as seeds and uses an LLM to generate natural-language rewrite rules, which are then fed back to the model to rewrite queries. These rules are reused for future queries. However, this approach requires per-query LLM interactions, struggles to scale to large workloads, and relies on verification by experts.
In contrast, ReSequel leverages metadata to guide template rewriting with caching and verification.

\begin{figure}[!t]
    \vspace{-0.35cm}
    \centering     
    \resizebox{.99\columnwidth}{!}{\includegraphics[scale=1]{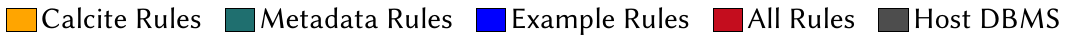}} \vspace{-0.17cm}    

    %\resizebox{.99\columnwidth}{!}{ 
    \subfigure[\footnotesize{Rules Ratio}]{\label{fig:ab-a}\includegraphics[scale=0.82]{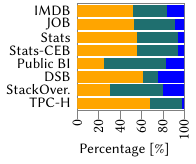}}
    \includegraphics[scale=0.42]{figures/img/fig_delimiter}
    \subfigure[\footnotesize{Overall (JOB)}]{\label{fig:ab-b}\includegraphics[scale=1.07]{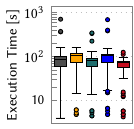}}
    \subfigure[\footnotesize{Individual (JOB)}]{\label{fig:ab-c}\includegraphics[scale=0.86]{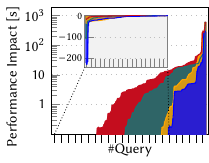}}
    %}
    \vspace{-0.6cm}  
    \caption{Rule Distribution Across Workloads and Relative Rule Impact on JOB (Gemini-2.5-pro, PostgreSQL).}    
    \vspace{-0.65cm}    
\end{figure}

\textbf{Rewrites in Query Optimization:} Beyond heuristic rewrite systems, prior work has integrated rewrites directly into cost-based query optimization. The Cascades framework~\cite{Graefe95a} applies rewrite rules during exploration of group expressions, intertwining rewrite application with cost-based join enumeration. 
Other approaches integrate specific rewrites into join optimization, such as pre-aggregation before joins~\cite{ChaudhuriS94,YanL95}. In addition, cost-based optimizers have been extended with specialized physical operators, including group-join~\cite{MoerkotteN11}, which captures common rewrite patterns combining joins and group-by aggregation on the same key.
%
% In contrast to these systems, we perform holistic rewrites over join orders and related transformations in an instance-centric setting, targeting individual queries.
In contrast, we perform holistic rewrites over join orders and related transformations on individual query instances.

% \textbf{LLM-based Code Generation:} LLMs are increasingly used to generate code artifacts, including information extraction programs~\cite{ShankarCSPW25}, data wrangling scripts~\cite{NarayanCOR22}, data analysis and feature engineering pipelines~\cite{Hollmann0H23}, and data-centric ML pipelines~\cite{FathollahzadehMB25}. Many of these tasks require specialized prompt engineering and additional metadata, particularly for non-public datasets that are not part of LLM training corpora. Beyond functional correctness, recent work also studies benchmarks and LLM-based techniques for improving program runtime efficiency~\cite{0005QSCZ24,0005DWWQCG024,huang2025efficoder}. These approaches generate and evaluate multiple candidate programs.
% %
% In contrast, ReSequel groups queries into templates, rewrites templates rather than individual queries, and selects the best candidates through execution-based verification on database samples.

\textbf{Automatic Query Verification:} Verifying rewritten queries is essential for correctness. Traditional systems define equivalence using source-target patterns~\cite{moerkotte25} and apply them iteratively. Automatic rewrite discovery systems also rely on verification~\cite{WeTune}. SQLSolver~\cite{DingWYZXCPL23} determines query equivalence using linear integer arithmetic and existing solvers. Beyond these methods, prior work on SQL testing~\cite{RiggerS20} and text-to-SQL generation~\cite{PapicchioPC23} compare query results directly. ReSequel follows this result-oriented approach, comparing outputs across rewritten templates on samples.

%% file: 9_Conclusions.tex
% \vspace{-0.35cm}
\section{Conclusions}

We introduced ReSequel, an end-to-end query rewriting system that operates on top of existing DBMSs and uses LLMs to optimize groups of related queries. ReSequel combines template-level rewriting with verification and Top-1 query selection.
We draw three conclusions. First, despite decades of work on query rewriting, large rule sets and their interactions remain brittle. Phase ordering, limited rewrite budgets, and strict pattern matching prevent DBMSs from fully optimizing messy queries. 
Second, rewriting recurring query templates, rather than individual queries, substantially speeds up rewriting. This approach enables scalable, reusable, and robust optimization.
Third, in a metadata- and statistics-guided manner, LLMs are capable of correct query rewriting that yields substantial runtime improvements on various DBMSs.
Future work includes generating DBMS-specific rewrite rules from verified LLM-generated rewrites for integration into existing DBMS code bases.